\documentclass[useAMS,usenatbib]{mn2e}
\usepackage{graphicx}
\usepackage{epsfig}

\newcommand{\kms}{\mbox{km s$^{-1}$} }

\newcommand{\Vmax}{\mbox{$V_{\rm max}$}}

\def\la{\mathrel{\hbox{\rlap{\hbox{\lower4pt\hbox{$\sim$}}}\hbox{$<$}}}}
\def\ga{\mathrel{\hbox{\rlap{\hbox{\lower4pt\hbox{$\sim$}}}\hbox{$>$}}}}

\newcommand{\bc}{\begin{center}}
\newcommand{\ec}{\end{center}}

\newcommand{\be}{\begin{equation}}
\newcommand{\ee}{\end{equation}}

\newcommand{\vechm}[1]{\overrightarrow{\bf #1} }

\newcommand{\tx}[1] {\rmn{#1}}

\newcommand{\MOPED} {\hbox{MOPED}}

\newcommand{\SN}{\rm{S/N}}
\newcommand{\SNmed}{{\rm S/N}}
\newcommand{\sigv}{\hbox{$\sigma_V$}}

\title[Extracting star formation histories from medium-resolution galaxy 
spectra]{Extracting star formation histories from medium-resolution galaxy spectra}
\author[H. Mathis, S. Charlot and J. Brinchmann]
	{H.~Mathis$^{1,2}$, 
	S.~Charlot$^{2,3}$\thanks{Email: charlot@iap.fr} and  J.~Brinchmann$^{2,4}$	
	\\	
        $^1$Department of Astrophysics, University of Oxford, Keble Road, OX1 3RH, UK \\
        $^2$Max--Planck--Institut f\"ur Astrophysik, Karl-Schwarzschild-Strasse 1, 85748 Garching, Germany \\        
        $^3$Institut d'Astrophysique de Paris, UMR7095 CNRS, Universit\'e Pierre \& Marie 
	Curie, 98 bis boulevard Arago, 75014 Paris, France \\
        $^4$Centro de Astrofisica da Universidade do Porto, Rua das Estrelas, 4150-762
Porto, Portugal \\
	}

\begin{document}

\date{MNRAS, in press}

\maketitle

\label{firstpage}



\begin{abstract}

We adapt an existing data compression algorithm, MOPED, to the extraction of 
median-likelihood star formation histories from medium-resolution galaxy spectra.
By focusing on the high-pass components of galaxy spectra, we minimize potential
uncertainties arising from the spectro-photometric calibration and intrinsic attenuation
by dust. We validate our approach using model high-pass spectra of galaxies with 
different star formation histories covering the wavelength range 3650--8500~{\AA} at a 
resolving power of $\sim2000$. We show that the method can recover the full star formation
histories of these models, without prior knowledge of the metallicity, to within an 
accuracy that depends sensitively on signal-to-noise ratio. The investigation of the 
sensitivity of the flux at each wavelength to the mass fraction of stars of different ages
allows us to identify new age-sensitive features in galaxy spectra. We also highlight a
fundamental limitation in the recovery of the star formation histories of galaxies for
which the optical signatures of intermediate-age stars are masked by those of younger and
older stars.

As an example of application, we use this method to derive average star formation histories
from the highest-quality spectra of typical (in terms of their stellar mass),
morphologically identified early- and late-type galaxies in the Early Data
Release (EDR) of the Sloan Digital Sky Survey (SDSS). We find that,  in agreement with the 
common expectation, early-type galaxies must have formed most of their stars over 8~Gyr ago,
although a small fraction of the total stellar mass of these galaxies may be accounted for 
by stars with ages down to 4~Gyr. In contrast, late-type galaxies appear to have formed 
stars at a roughly constant rate. We also investigate the constraints set by the high-pass
signal in the stacked spectra of a magnitude-limited sample of 20,623 SDSS-EDR galaxies 
on the global star formation history of the Universe and its distribution among
galaxies in different mass ranges. We confirm that the stellar populations in the most 
massive galaxies today appear to have formed on average earlier than those in the least
massive ones. Our results do not support the recent suggestion of a statistically 
significant peak in the star formation activity of the Universe at redshifts below unity,
although such a peak is not ruled out.

\end{abstract}




\section[]{Introduction}
\label{sec:Intro}

Modern spectroscopic galaxy surveys such as the 2dF Galaxy Redhift Survey \citep{Colless01} 
and the Sloan Digital Sky Survey (SDSS, \citealt{Stou02}) are collecting hundreds of 
thousands of medium-resolution spectra of galaxies in the nearby universe. At higher 
redshift, the VIRMOS-VLT Deep Survey \citep{LeF01} and the Deep Extragalactic Evolutionary
Probe \citep{Dav03} are collecting the spectra of nearly 200\,000 galaxies in the younger 
universe. The high-quality spectra gathered by these surveys exhibit a myriad of stellar 
absorption features containing valuable information about the physical properties of 
the galaxies, such as age, star formation history, metallicity and dust content. The
extraction of this information from medium-resolution galaxy spectra is essential for
improving current constraints on galaxy evolution.

Stellar population synthesis, the modeling of the light emitted by specific populations of
stars, is a natural approach to interpreting observed galaxy spectra in terms of physical
parameters. Until recently, the spectral resolution of standard population synthesis codes
was typically 3--10 times lower than that achieved in modern spectroscopic galaxy surveys
($\lambda/\Delta \lambda \sim2000$ for the SDSS). Thus, for spectral analyses to be 
performed, observed high-quality galaxy spectra had to be {\em degraded} to the resolution
of the models, resulting in the loss of valuable high-frequency information. The advent of
new models with higher spectral resolution has opened the door to more refined spectral
analyses (\citealt{Vaz99}; \citealt{Bru03}). For example, the ability to model in detail the
4000~{\AA} discontinuities and H$\delta_{\rm A}$ absorption indices of more than $10^5$ 
SDSS galaxies allowed \citet{Ka03a} to derive useful constraints on the mass fraction of stars
formed in recent bursts in these galaxies. The next step is to identify the information
provided by the {\em entire} medium-resolution spectrum on the star formation history,
metallicity and dust content of a galaxy. 

The extraction of physical parameters from large numbers of medium-resolution galaxy 
spectra requires the development of dedicated techniques. Several algorithms have been 
designed to efficiently `compress' galaxy spectra into a reduced number of parameters with
minimum loss of information. Such algorithms include, for example, principal component 
analysis \citep{Ro99}, Fisher matrix analysis \citep{Teg97} and information bottleneck (see
\citealt{La01} for a review). Some of these algorithms have also been used to 
classify galaxy spectra, complementing pure classification methods such as trained neural
networks. So far, however, these methods have not been used to extract physical parameters
from {\em medium-resolution} galaxy spectra. 

In this paper, we use an original approach to extract star formation histories (and rough
constraints on the metallicities) from medium-resolution galaxy spectra. Our approach relies
on the existing algorithm \MOPED\footnote{The \MOPED\ algorithm has a patent.} (Multiple 
Optimised Parameter Estimation and Data compression) developed by \citet{Hea00}, which 
allows one to compress galaxy spectra into a reduced number of linear combinations connected
to `physical parameters' (for example, the mass fraction of stars in a given age range). 
\MOPED\ has been used in the past to recover star formation histories, metallicities and 
dust contents from low-resolution galaxy spectra \citep{Reic01, Pan03}. The originality of
our work consists in combining \MOPED\ with a recent stellar population synthesis model of
medium resolution \citep[hereafter BC03]{Bru03}. We focus on the high-pass components of 
galaxy spectra, which allows us to minimize potential uncertainties arising from the 
spectro-photometric calibration and from the  intrinsic attenuation by dust. This requires us
first to remove any contamination of the high-frequency signal by nebular emission lines. 
Our approach also differs from previous applications of the MOPED\ algorithm in that we 
compute full likelihood distributions of physical parameters and do not rely on acceleration
techniques to selectively explore the parameter space. This allows us to construct more 
realistic confidence intervals for the derived star formation histories than has been 
previously possible.	

We show that our method succeeds in recovering the star formation histories from a variety
of model galaxy spectra, to an accuracy that increases with signal-to-noise ratio. We 
investigate the spectral features that are most sensitive to stars in various age ranges,
excluding those regions of the optical spectrum that are known to depend strongly on 
changes in metal abundance ratios. We also point out a fundamental limitation in the spectral
interpretation of galaxies for which the optical signatures of intermediate-age stars are
masked by those of younger and older stars. To illustrate the usefulness of our approach,
we use it to interpret the spectra of galaxies with high signal-to-noise ratios in 
the SDSS early data release (EDR). We find that, in general, galaxies with high 
concentrations typical of early-type galaxies appear to have formed the majority of their
stars prior to $z=1$, while late-type, low-concentration galaxies appear to have formed 
stars at a roughly constant rate at all times. From the analysis of the stacked spectra 
of a magnitude-limited sample of 20,623 SDSS-EDR galaxies, we can also investigate
the global cosmic star formation history of the Universe. Our results confirm previous
evidence that the stars that end up in massive galaxies today formed on average earlier 
than those that end up in low-mass galaxies (\citealt{Hea04}; see also \citealt{Cow96};
\citealt{Kod04,Kau04}). Our analysis does not reveal any statistically significant peak in
the integrated star formation activity of the Universe at redshifts less than $z\sim1$.

The paper is organized as follows. In Section~\ref{sec:Method}, we describe our approach
and the population synthesis models we rely on. We also recall the main features of the 
\MOPED\ algorithm. In Section~\ref{sec:Testing}, we use model simulations to evaluate the 
extent to which star formation histories can be recovered from medium-resolution galaxy
spectra. We also investigate the spectral features that are most revealing of stars in 
various age ranges. In Section~\ref{sec:EDR}, we first derive star formation histories from
individual, high-quality spectra of high- and low-concentration galaxies in the SDSS EDR.
We also investigate the typical star formation histories of high- and low-concentration
galaxies, based on the analysis of stacked spectra. We complement this with a study of the
overall star formation history of the Universe using a sample of 20,623 galaxies from the
SDSS EDR. Our conclusions are summarized in Section~\ref{sec:CCL}. Throughout this paper, 
we adopt a Friedmann-Robertson-Walker cosmology with $\Omega=0.3$, $\Lambda=0.7$ and 
$H_0=70\,\rm km\, s^{-1}\,Mpc^{-1}$.



\section[]{The method}
\label{sec:Method}

We summarize here our approach for extracting physical parameters from medium-resolution
galaxy spectra. We first briefly review the main features of the compression algorithm 
\MOPED\ and how we adapt it to interpret the high-pass signal in galaxy
spectra. We describe the physical parameters of interest to us and 
 how we estimate their likelihood. 

\subsection[]{The \MOPED\ algorithm}
\label{sec:Method:Overview}
	
MOPED is a linear compression algorithm, which efficiently reduces a large array of data
points into an arbitrarily small number of key parameters. Such an algorithm is ideally 
suited to the extraction of physical parameters from medium-resolution galaxy spectra. 
We refer the reader to the original paper of \citet[][see also \citealt{Teg97}]{Hea00}
for a more detailed description than given in the following summary.

\begin{itemize}
\item In \MOPED, the compression of a galaxy spectrum $\vechm{F}_{\lambda}$ (defined as a
vector of fluxes per unit wavelength) into a small number ${n}_{\rmn{P}}$ of physical
parameters -- for example, the mass fraction of stars in a given age range -- is 
represented by a vector $\vechm{y}$ of ${n}_{\rmn{P}}$ elements $y_j$, defined by $y_j=
\vechm{b}_{j} \cdot \vechm{F}_{\lambda}$. 

\item The ${n}_{\rmn{P}}$ weighting vectors $\vechm{b}_j$ are specific to the spectrum
$\vechm{F}_{\lambda}$. They quantify in an optimal way the sensitivity of the flux at 
each wavelength to variations in the $j^{\rm th}$ parameter, taking into account the noise
associated to $\vechm{F}_{\lambda}$. The construction of the weighting vectors
$\vechm{b}_j$ requires a population synthesis model to compute the derivatives of the flux
at each wavelength with respect to each physical parameter. In practice, for the sake of
efficiency, these derivatives are estimated at a single point in the parameter space called
the `fiducial model' (see Section~\ref{sec:Method:Like:Fiducial} below). The resulting
vectors $\vechm{b}_j$ contain information about the spectral features most sensitive to the
parameter $j$, given the noise pattern.

\item Once the weighting vectors $\vechm{b}_j$ are determined, the probability distributions
of the ${n}_{\rmn{P}}$ parameters given the spectrum $\vechm{F}_{\lambda}$ may be evaluated
by computing the likelihood of each (compressed) model $\vechm{y}_{\rmn{mod}}$ in the 
parameter space, according to the norm of the vector $\vechm{y}-\vechm{y}_{\rmn{mod}}$.

\end{itemize} 

The \MOPED\ compression would not cause any loss of information about the ${n}_{\rmn{P}}$ 
parameters if these had Gaussian likelihood distributions, and if the fiducial model 
corresponded for each $\vechm{F}_{\lambda}$ to the best fit model that would be obtained
without compression. In most applications however, these criteria are not met 
(Section~\ref{sec:Method:Like:Sampl}). It is also worth noting that, because the weighting
vectors $\vechm{b}_j$ are obtained recursively in a process similar to a Gram-Schmidt 
orthonormalization of independent vectors, the values of the weights may depend on the
order in which the parameters are considered. This is especially true when some spectral
features present similar sensitivities to several parameters (see also
Section~\ref{sec:Method:ParSpace:Z}). We have found that, in general, our results are 
relatively robust to changes in the order in which the weighting vectors $\vechm{b}_j$ are evaluated. We therefore 
 adopt  here the natural order used by \citet{Reic01} and \citet{Pan03}, which consists in evaluating the sensitivity of the
spectrum to the oldest stars first, and then proceed with younger stars.

As mentioned in Section~\ref{sec:Intro}, the \MOPED\ algorithm has already been used to 
extract star formation histories from low-resolution spectra of galaxies in the Kennicutt
(1992) atlas and the SDSS survey \citep{Reic01, Pan03}. In the remainder of this section,
we describe our application of this algorithm to the interpretation of medium-resolution
galaxy spectra.

\subsection[]{Medium-resolution galaxy spectra}
\label{sec:Method:Spec:Cover}

To interpret medium-resolution galaxy spectra with \MOPED, we rely on the recent population
synthesis code of BC03. This is based on the \citet{LeB03}  library of observed stellar 
spectra and has a resolution of 3~{\AA} across the whole wavelength range from 3200~{\AA} 
to 9500~{\AA} (corresponding to a median resolving power $\lambda/\Delta \lambda \approx
2000$ and a nominal velocity dispersion of 70 $\kms$), for a wide range of metallicities
($-2.0 <{\rm [Fe/H]} < +0.50$). Since we are interested here in the interpretation
of SDSS galaxy spectra (see Section~\ref{sec:EDR}), we restrict our analysis to the 
wavelength range from 3650 to 8500~{\AA}, where the noise is lowest in the rest-frame SDSS
spectra. To speed up the analysis, we increase the wavelength sampling of the BC03 model 
spectra from the original 1-{\AA} sampling to a 3-{\AA} sampling. Furthermore, in all 
applications below, we always use model spectra broadened to a velocity dispersion
consistent with that of the galaxy under study. 

The observed stellar spectra incorporated in the BC03 models come from nearby stars with
solar abundance ratios. However, stars in external galaxies are known to exhibit 
changes in heavy-element abundance ratios. In particular, the ratio of $\alpha$ elements to
iron appears to be enhanced in massive early-type galaxies relative to the solar composition
(e.g., \citealt{Wor92,Eis03}). To increase the applicability of our models to the 
interpretation of observed galaxy spectra, we choose to mask out those regions of the
optical spectrum that are known to strongly depend on changes in metal abundance ratios.

BC03 used a large library of model star formation histories to fit strong atomic and
molecular features in the highest-quality spectra of the SDSS EDR. They considered the 25
commonly measured spectral features of the `Lick system' that were defined and 
calibrated in the spectra of 460 Galactic stars in the wavelength range from 4000~{\AA} to
6400~{\AA} \citep{Wor94,Wor97,Tra98}. As expected, the Lick features that are the least 
well fitted by the BC03 models in the SDSS spectra are those \citet{Th03a} found to depend
most strongly on changes in $\alpha/\tx{Fe}$.  We therefore exclude from our analysis the 
spectral bandpasses corresponding to the Lick features found by \citet{Th03a} to be most 
sensitive to changes in $\alpha/\tx{Fe}$. These are Mg$_{1}$, Mg$_{2}$, Fe4383, Fe5335 and 
Fe5406. The wide TiO$_{1}$ and TiO$_{2}$ features are also not always well reproduced by the
BC03 models in the highest-quality SDSS spectra. However, we keep them in the present 
analysis because they are thought to depend less strongly on $\alpha/\tx{Fe}$ \citep{Th03a}.
We also do not exclude the spectral bandpasses defining other Lick indices, which sometimes
overlap with the bandpasses of Mg$_{1}$, Mg$_{2}$, Fe4383, Fe5335 and Fe5406. In summary, 
the spectral windows we exclude from our analysis are 4363--4424\,\AA, 4445--4458\,\AA, 
4897--4958\,\AA, 5071--5135\,\AA, 5157--5199\,\AA, 5318--5367\,\AA\ and 5377--5425\,\AA, 
corresponding to a total spectral width of 338\,\AA. 

Our main motivation in the present study is to identify the constraints set by the
high-pass signal\footnote{In this paper, we use the term `high pass' to designate that
part of a galaxy spectrum containing high-frequency information. We prefer the term `high
pass' over `high frequency' because we do not strictly perform true frequency
filtering.} in galaxy spectra on the past history of star formation. We choose not 
to include the low-pass components of galaxy spectra for several reasons.
Firstly, the low-pass signal is strongly affected by dust attenuation and potential 
spectro-photometric inaccuracies, which introduce unwanted uncertainties in the derived
star formation histories. Secondly, population synthesis models with medium spectral 
resolution have only recently become available, and it is important to establish the 
extent to which the high-pass signal helps us constrain the physical parameters of 
galaxies. Finally, the \MOPED\ algorithm has already been used to interpret low-resolution
galaxy spectra in terms of star formation histories (see Section~\ref{sec:Method:Overview}).

We compute the high-pass component $F_{\lambda}^{\rmn{HP}}$ of a galaxy spectrum $F_\lambda$
(either modelled or observed) by first smoothing $F_\lambda$ with a top-hat kernel 
$W(\lambda)$ of smoothing diameter $l_{\rmn{s}}=500\;$\AA. This yields the low-pass spectral
component
\be
F_{\lambda}^{\rmn{LP}}=\int^{\lambda+l_{\rmn{s}}/2}_{\lambda-l_{\rmn{s}}/2} \,
d\lambda^\prime\,W(\lambda-\lambda^\prime)\,F_{\lambda^\prime}\,.
\label{eqn:LP}
\ee
The high-pass component may then be obtained as either 
\be
F_{\lambda}^{\rmn{HP}}=\frac{\displaystyle{F_\lambda}}
		      {\displaystyle{F_{\lambda}^{\rmn{LP}}}}
\label{eqn:HPfromD}
\ee
or
\be
F_{\lambda}^{\rmn{HP}}= F_\lambda-F_{\lambda}^{\rmn{LP}}\,.
\label{eqn:HPfromS}
\ee
Equation\,(\ref{eqn:HPfromD}), which was used by \citet{Bal02}, preserves the equivalent
widths of individual spectral features. There is no strong {\em a priori} motivation for 
using equations~(\ref{eqn:HPfromD}) or (\ref{eqn:HPfromS}) to compute the high-pass 
components of galaxy spectra. Equation~(\ref{eqn:HPfromS}), which would be preferred if we 
achieved true frequency filtering, appears to perform better in the recovery of star 
formation histories in model simulations. We therefore use it throughout this paper
to compute high-pass galaxy spectra (we have checked that using equation~\ref{eqn:HPfromD}
would not affect our main conclusions). Since the noise associated to the low-pass component 
is negligible relative to that associated to the high-pass component, the signal-to-noise
ratio per pixel in the high-pass spectrum is virtually the same as that in the original
spectrum. 

The high-pass galaxy spectra obtained using equation~(\ref{eqn:HPfromS}) must be normalized
for the purpose of comparisons between models and observations. Ideally, the normalization
should be achieved in a band which traces the total stellar mass of the galaxy. In practice,
we are limited by the wavelength coverage and redshifts of the SDSS spectra we plan to 
analyze. We have found that the rest-frame 6000--7300 {\AA} band provides a good compromise
and therefore normalize all spectra below in this wavelength range. All results presented
here are based on a \citet{Cha03} initial mass function with lower and upper cutoffs 0.1 and
100\,$M_\odot$.

\subsection[]{Physical parameters}
\label{sec:Method:ParSpace}

We now describe the physical parameters we wish to recover from medium-resolution galaxy
spectra. We are primarily interested in the past history of star formation. We also derive
constraints on the stellar metallicity and mention the potential of our approach for 
constraining the attenuation of starlight by dust in galaxies.

\subsubsection[]{Star formation history}
\label{sec:Method:ParSpace:PopAges}

We characterize the star formation history of a galaxy by constraining the evolution of the
star formation rate in 6 consecutive age bins. Ideally, these age bins should correspond to
significant changes in the spectral characteristics of the stars, such as those arising
through the onset of red supergiants, bright asymptotic giant stars, late-B and A stars 
with strong H-Balmer absorption lines, and red giant branch stars of different 
temperatures.\footnote{\citet{Reic01} and \citet{Pan03} use 8 age bins to describe the star
formation history of a galaxy in their analysis of low-resolution spectra using \MOPED. 
While the computationally more demanding interpretation of {\em medium-resolution} spectra
prevents us from using such a large number of bins, we find that adopting more than  6 bins
does not statistically improve the constraints we derive on the star formation history.
This is because the spectral signatures of stars of too similar ages cannot be easily 
distinguished, even at medium resolution.} After some experimentation, we settled on the 
following age bins to characterize the star formation history of a galaxy over the period 
from the age of the Universe to the present time: [13.5--8], [8--4], [4--1.5], [1.5--0.5],
[0.5--0.1] and [0.1--0]~Gyr. Within each bin, star formation is assumed to proceed at a
constant rate. The relative `amplitude' of a bin then determines the fraction of the total
stellar mass of a galaxy that formed at the epoch corresponding to that bin.

The sampling of galaxy star formation histories in the \MOPED\ algorithm is defined by the
possible combinations of star formation amplitudes in the above 6 age bins. We have chosen
6 levels of star formation amplitude in each age bin, such that the {\em fraction} of the 
total stellar mass of a galaxy formed in any age bin be sampled in an optimal way among all 
possible star formation histories. For each of the 3 oldest age bins, we require that this 
fraction be linearly well sampled (i.e. with approximately equal numbers of models) by 6 
values between 0 and 1. For each of the 3 youngest age bins, we require that it be 
logarithmically well sampled by 6 values between $10^{-3}$ and 1 (we do not distinguish 
fractions of $10^{-3}$ or less from zero star formation). To select the absolute star formation
amplitudes providing the closest approximation to such a sampling in stellar mass fraction, we
have used the non-negative least squares routine of \citet{Law74}. We find that 6 amplitudes
per bin are sufficient to sample the parameter space in an adequate way. Adopting a 
computationally more expensive sampling of 8 amplitudes per bin does not significantly improve
the accuracy to which the method can recover star formation histories from medium-resolution
galaxy spectra (Section~\ref{sec:Testing:MedianTime}).

\subsubsection[]{Metallicity}
\label{sec:Method:ParSpace:Z}

For the purpose of the present paper, we assume that stars of all ages in a given galaxy
have the same metallicity. We interpret this as the `effective' stellar metallicity (i.e.,
the metallicity of the stars dominating the light). This choice is motivated primarily 
by the fact that our models have solar heavy-element abundance ratios, while the abundance
ratios of heavy elements are known to vary in external galaxies (e.g., \citealt{Wor92,Eis03}). 
As mentioned in Section~\ref{sec:Method:Spec:Cover}, we exclude from our
analysis the spectral features with the strongest identified dependence on changes in the
$\alpha$/Fe abundance ratio.  While we believe that this is sufficient for obtaining 
rough constraints on effective metallicities, we are less confident in the reliability
of the models for recovering accurate metal enrichment histories from observed galaxy 
spectra. This caution is further justified by the fact that the signatures of metals are 
often similar in the spectra of stars with different ages. Allowing different metallicities
for different age bins would therefore introduce an unwanted dependence of the results on 
the order in which the \MOPED\ weighting vectors are computed (see 
Section~\ref{sec:Method:Overview}). We note that the star formation histories derived in 
Sections~\ref{sec:Testing} and \ref{sec:EDR} below are marginalized over effective 
metallicity (Section~\ref{sec:Method:Like:Sampl}).

For simplicity, we consider here only 3 choices for the effective stellar metallicity of 
a galaxy: $Z=0.4Z_\odot$, $Z_\odot$ and $1.5Z_\odot$. These values are representative of
the range in stellar metallicities inferred from the analysis of selected features with
negligible dependence on $\alpha$/Fe in the spectra of $\sim2\times10^5$ SDSS galaxies
\citep{Galla05}.

\subsubsection[]{Attenuation by dust}
\label{sec:Method:ParSpace:Dust}

The analysis presented in this paper, which relies on the interpretation of high-pass
optical galaxy spectra, is largely exempt from the uncertainties arising from attenuation
of starlight by dust in galaxies (Section~\ref{sec:Method:Spec:Cover}). The potential weak
effects of dust on the relative strengths of stellar absorption features are expected to 
affect only negligibly the star formation histories derived from high-pass galaxy spectra
(see Section~\ref{sec:Testing:MedianTime}).

It is worth mentioning that our approach also provides a means of constraining attenuation 
by dust. The star formation history (and effective metallicity) derived from the high-pass
spectrum of a galaxy can be exploited to predict the low-pass spectrum of that galaxy using
dust-free population synthesis models. By comparing this low-pass spectrum with the
observed one, it is possible to derive useful constraints on the attenuation by dust in
the galaxy (for example, using the model of \citealt{Ch00}). Unlike the
analysis of the high-pass spectrum, this requires an accurate spectro-photometric 
calibration of the observed spectrum.

\subsection[]{Statistical estimates of physical parameters}
\label{sec:Method:Like}

One of the main advantages of the \MOPED\ algorithm over other  techniques to interpret
galaxy spectra is that it allows the derivation of {\em statistical} estimates of physical
parameters. We outline below the way in which this is achieved in our models.

\subsubsection[]{Choice of the fiducial model}
\label{sec:Method:Like:Fiducial}

In the \MOPED\ algorithm, the fiducial model is the point of the parameter space where
the weighting vectors describing the sensitivity of the spectrum to the various physical 
parameters are evaluated (Section~\ref{sec:Method:Overview}). The vectors are constructed
by computing the derivatives of the flux at each wavelength with respect to each parameter 
 and accounting for the noise pattern.

The requirement in choosing the fiducial model is that its spectrum should contain
as many spectral features as possible to constrain the different parameters of interest. 
To constrain the star formation history, the spectrum should therefore have clear signatures
of stars in all age ranges. In the present study, the choice of the fiducial model is 
simplified by the fact that we are interested only in the high-pass components of galaxy 
spectra, as obtained by subtracting the low-pass components from the total spectra 
(equation~\ref{eqn:HPfromS}). Also, in our models, galaxy spectra are simple linear 
combinations of the spectra of stellar populations of identical metallicity in 6 age bins
(Section~\ref{sec:Method:ParSpace:PopAges}). Finally, we neglect any potential weak, 
non-linear effect of dust on high-pass galaxy spectra. All this implies that, at fixed 
metallicity and stellar velocity dispersion, the derivatives of the high-pass component 
with respect to physical parameters do not depend on the specific model (as characterized
by the amplitudes of the stellar populations in the 6 age bins) at which they are evaluated,
so long as this model includes contributions from all age bins.\footnote{It is interesting 
to note that this would not be the case if we had chosen equation~(\ref{eqn:HPfromD}) instead
of equation~(\ref{eqn:HPfromS}) to compute high-pass galaxy spectra. In this case, the 
low-pass component would enter the expression of the derivatives of the high-pass component
with respect to the physical parameters.} We compute such a set of high-pass
derivatives for each of the 3 metallicities considered here, $Z=0.4Z_\odot$, $Z_\odot$ and
$1.5Z_\odot$, and for several stellar velocity dispersions between 70 and 300\,\kms 
(Sections~\ref{sec:Method:Spec:Cover} and \ref{sec:Method:ParSpace:Z}).

We take the noise spectrum entering the computation of the \MOPED\ weighting vectors
at the fiducial model to be that of the observed galaxy under study. The disadvantage is 
that this requires us to evaluate the weighting vectors for every galaxy in the observed 
sample.\footnote{\citet{Reic01} and \citet{Pan03} adopt a fixed, `average' noise spectrum
to interpret low-resolution galaxy spectra with \MOPED.} The reason for this refinment is to
limit the loss of information about the physical parameters arising from the compression of
the spectra. As emphasized by \citet{Hea00}, the \MOPED\ compression is optimal when the
properties of the fiducial model are as close as possible to those of the observed spectrum.

\subsubsection[]{Probability density functions}
\label{sec:Method:Like:Sampl}

The likelihood that any model in the parameter space represents a given observed spectrum
may be computed by comparing the compressed model and observed spectra as described in 
Section~\ref{sec:Method:Overview}. If the likelihood distributions of the physical 
parameters were well approximated by multivariate Gaussian functions over the entire 
parameter space, sampling the neighbourhood of the most likely model would be sufficient
to estimate the confidence intervals for these parameters. In our analysis, we have found
that the likelihood distributions of the fractions of stars formed in different age bins
is not necessarily well approximated by Gaussian functions. Thus, the parameter space must
be explored more thoroughly to derive constraints on the star formation histories of galaxies.

Sampling the entire parameter space to evaluate the likelihood of every model given an 
observed spectrum is computationally demanding. For this reason, previous studies appealed
to standard techniques such as `Monte-Carlo Markov Chains' to selectively explore the
parameter space and construct the likelihood distributions of physical parameters
(\citealt{Pan03}; see also \citealt{Ve03}). In practice, such techniques rely on the 
sampling of the parameter space by a fixed, limited number of points. As a consequence, 
it is unclear whether the parameter space is always sampled in an adequate way to estimate
the true probability density functions of physical parameters.

We choose here to favor the accuracy of error estimates over the reduction of computational
cost and always explore the entire parameter space to compute the likelihood distributions
of the different physical parameters. We first compute, for each metallicity, the Bayesian
likelihood distribution of the fraction of stars in each age bin. This is evaluated by 
weighting each model by its probability  and then summing the probabilities in bins 
of fraction of stars (see appendix~A of \citealt{Ka03a}). We then marginalize the 
likelihood distribution of the fraction of stars in each age bin over metallicity, by
summing the likelihood distributions over  the 3 metallicities $Z=0.4Z_\odot$, $Z_\odot$
and $1.5Z_\odot$. In  this procedure, we assume that the 3 metallicities are equally
probable a priori. We refer to the result of this sum as the probability density function
of the fraction of stars in that age bin. By construction, this function includes the errors
associated to the metallicity determination. We also obtain a rough estimate of the most 
likely metallicity of the galaxy by marginalizing the likelihood distributions obtained for
each metallicity over star formation histories.

\subsubsection[]{Priors}
\label{sec:Method:Like:Priors}

In our models, the prior distributions of the parameters are flat, reflecting a point of
view of `maximum ignorance'. The fraction of stellar mass formed is uniformly distributed 
in each of the oldest 3 bins (ages $\ge1.5\,$Gyr) and logarithmically distributed in each 
of the youngest 3 bins (ages $<1.5\,$Gyr; see Section~\ref{sec:Method:ParSpace:PopAges}). 
In addition, the models are uniformly distributed in metallicity between $Z=0.4Z_\odot$,
$Z_\odot$ and $1.5Z_\odot$ (Sections~\ref{sec:Method:ParSpace:Z} and 
\ref{sec:Method:Like:Sampl}). We find that our results are  robust to 
changes in these prior distributions. For example, changing from a uniform 
to a logarithmic sampling of the stellar mass fractions (and conversely) has 
little effect on the derived star formation
histories. Such an invariance is expected when the problem is well constrained.

\section[]{Assessment of efficiency using model galaxy spectra}
\label{sec:Testing}

\subsection[]{Outline}
\label{sec:Testing:Intro}

This Section is the core of the paper. Here, we use model simulations to assess 
the accuracy to which our method is able to recover star formation histories from 
medium-resolution galaxy spectra. For illustration, we parametrize the star
formation history as a simple exponential law $\rmn{SFR}(t)\propto \exp{(-t/\tau)}$, where
$t$ is the age of the galaxy and $\tau$ the time scale of star formation. We consider 3 
models with $\tau=2,$ 7~Gyr and $\infty$ (i.e. constant star formation rate), all with the
same present galaxy age $t=13.5\,$Gyr (correponding to the age of the Universe for the
adopted cosmology). These simple models are often taken to represent the star formation 
histories of, respectively, E/S0, Sb/Sc and Sm/Im galaxies (e.g., \citealt{Ka98}). For 
simplicity, we assume that all 3 models have solar metallicity and a stellar velocity 
dispersion $\sigv=100\,$\kms (we discuss below the effects of changing these parameters). 
We consider two values of the median signal-to-noise ratio per pixel in the spectra, 
$\SNmed=10$ and $\SNmed=30$. To be representative, the results presented here are always
averaged over 50 independent realizations of the noise (increasing this number from 50 
to 100 has a negligible effect on the results).

In Section~\ref{sec:Testing:MedianTime} below, we evaluate how accurately our method can
recover the star formation histories of these models, based on the analysis of high-pass 
spectra. Then, in Section~\ref{sec:Testing:Signatures}, we point out a fundamental limitation
in the recovery of the star formation histories of galaxies for which the optical signatures of 
intermediate-age stars are masked by those of younger and older stars. Finally, in 
Section~\ref{sec:Testing:NewInd}, we highlight the spectral features that are the most 
sensitive to the presence of stars in various age ranges.

\subsection[]{Recovery of star formation histories}
\label{sec:Testing:MedianTime}

We first consider the case of a model galaxy with a star formation time scale $\tau=7\,$Gyr.
This model is particularly interesting, because it corresponds to the expected time scale 
of star formation of the Universe when averaged over large volumes (e.g., \citealt{Glaz03};
\citealt{Brinch04}). We adopt for the moment $\SNmed=30$ and take the wavelength 
dependence of the noise to be that observed for a typical late-type galaxy in the SDSS
(Section~\ref{sec:EDR:DefGals}).

\begin{figure}
\begin{minipage}{8.5cm}
\epsfig{file=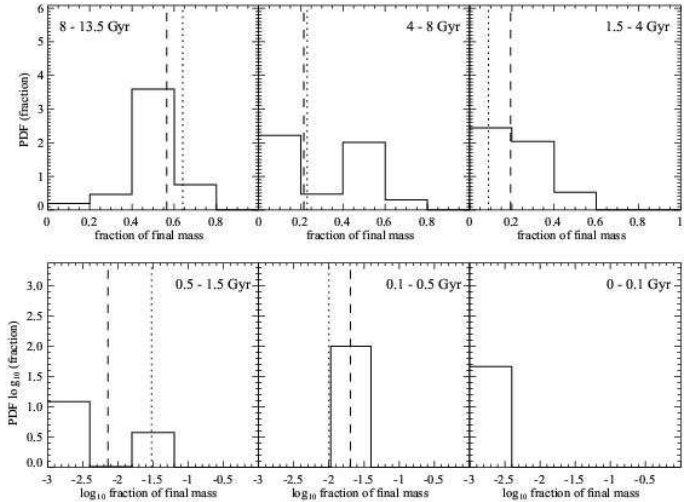,width=9cm} 
\caption{Test recovery of the star formation history from the high-pass spectrum of a 
13.5~Gyr old stellar population with a star-formation time scale of 7~Gyr and the solar
metallicity. In each panel, the histogram shows the average probability density function 
of the mass fraction of stars formed in a given bin of look-back time (as indicated).
This was obtained by summing the probability density functions obtained for 50 different
realizations of the noise. The dotted line shows the true fraction of stellar mass formed
in the original model, while the dashed line shows the most likely model recovered by the
analysis. The median signal-to-noise per pixel in the input spectrum is $\SN=30$, and the
stellar velocity dispersion $\sigv=100\,$\kms.}
\label{fig:Mock7GyrOldFormat}
\end{minipage}
\end{figure}

\begin{figure}
\begin{minipage}{8.5cm}
\epsfig{file=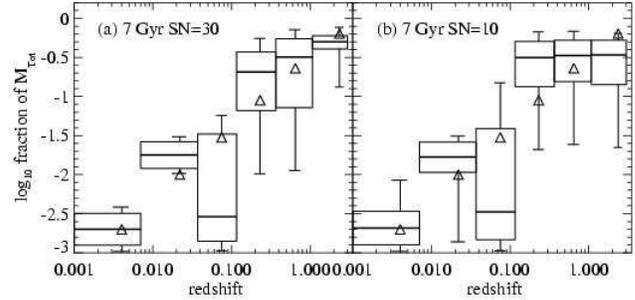,width=9cm}
\caption{(a) Same as Fig.~\ref{fig:Mock7GyrOldFormat}, but shown in a more compact way. In
each bin of stellar age, the thick line shows the median fraction of stars predicted to have
formed, based on the analysis of the model high-pass spectrum, while the vertical size of
the box and the error bars show the 16--84 and 2.5--97.5 percentile ranges in this fraction,
respectively (these ranges indicate the 68 and 95 percent confidence intervals measured from
the average probability density functions of Fig.~\ref{fig:Mock7GyrOldFormat}; they would 
correspond to the $\pm1\sigma$ and $\pm2\sigma$ confidence intervals in the case of Gaussian
distributions). The triangles indicate the true fractions of stellar mass formed in the model.
(b) Same as (a), but adopting a median signal-to-noise per pixel of 10 instead of 30 in the
input spectrum.}
\label{fig:Mock7Gyr}
\end{minipage}
\end{figure}

Fig.~\ref{fig:Mock7GyrOldFormat} shows the star formation history recovered from the
high-pass spectrum of this galaxy using the method presented in Section~\ref{sec:Method}.
Each panel shows the average probability density function of the mass fraction of stars 
formed in a given bin of look-back time. This was obtained by summing the 50 probability 
density functions obtained for different realizations of the noise (see above). The results 
are shown in a more compact way in Fig.~\ref{fig:Mock7Gyr}a, where we summarize the probability
distribution in each age bin by the median value (thick line) and 16--84 (box) and 2.5--97.5
(error bars) percentile ranges. We stress that these ranges do not reflect the scatter in 
the median mass fractions recovered from 50 different realizations of the noise. Instead, for
each age bin, the boxes and error bars in Fig.~\ref{fig:Mock7Gyr}a indicate the 68 and 95 
percent confidence intervals measured from the corresponding average probability density 
function of Fig.~\ref{fig:Mock7GyrOldFormat}. These ranges would correspond to the $\pm1\sigma$ 
and $\pm2\sigma$ confidence intervals in the case of Gaussian distributions. For reference,
the dotted lines in Fig.~\ref{fig:Mock7GyrOldFormat} and the triangles in 
Fig.~\ref{fig:Mock7Gyr}a show the true fractions of stellar mass formed at the different 
look-back times in the input model. We recall that the probability distributions in 
Figs~\ref{fig:Mock7GyrOldFormat} and \ref{fig:Mock7Gyr}a include the uncertainties associated
to the assumed unkown metallicity of the stars (Section~\ref{sec:Method:Like:Sampl}). The 
solar metallicity of the model is recovered with 70 percent confidence.

The results of Figs~\ref{fig:Mock7GyrOldFormat} and \ref{fig:Mock7Gyr}a indicate that,
overall, the method recovers remarkably well the star formation history of the test galaxy.
The true mass fraction of stars formed at different look-back times is always recovered
within the 16--84 percent ($\pm1\sigma$) confidence range predicted by the analysis.
Some discrepancies exist between the recovered and true star formation 
histories. In particular, the mass fraction of stars with ages around 1~Gyr appears not 
to be well constrained. Its median value is underestimated, while the median mass
fractions of stars in the adjacent age bins are overestimated. In 
Section~\ref{sec:Testing:Signatures} below, we show that this mismatch arises from a general
limitation of the spectral interpretation of galaxies for which the optical signatures of 
intermediate-age stars (0.5--4~Gyr) are masked by those of younger and older stars. 

The accuracy to which the star formation history can be recovered from high-pass galaxy
spectra depends sensitively on the signal-to-noise ratio. Fig.~\ref{fig:Mock7Gyr}b shows 
the same results as in Fig.~\ref{fig:Mock7Gyr}a, but assuming a median signal-to-noise ratio
per pixel of 10 instead of 30 in the input spectrum. In this case, the agreement between the
predicted and  the true star formation histories is significantly worsened. For example, the 
method cannot recover the differences in the relative fractions of stars in the oldest 3
bins. For $\SNmed=10$, the results also become sensitive to the order chosen to construct
the weighting vectors in the \MOPED\ algorithm (Section~\ref{sec:Method}).  

The effect of increasing the velocity dispersion at fixed $\SNmed$ is expected to be similar
to that of decreasing $\SNmed$ at fixed velocity dispersion. In practice, the amplitude of
the effect depends on the type of stars dominating the light. For example, changing the 
velocity dispersion from our adopted 100\,\kms to 300\,\kms has only a weak effect on the 
results presented in Figs~\ref{fig:Mock7Gyr}a and b. 

We have also checked that attenuation by dust affects negligibly the results presented
in Fig.~\ref{fig:Mock7Gyr}. We used the simple but physically motivated model of 
\citet{Ch00} to include the effects of attenuation by dust on the spectrum of the
$\tau=7\,$Gyr model galaxy. We adopted a typical $V$-band effective absorption optical depth 
$\hat{\tau}_V =1$ appropriate for SDSS galaxies (\citealt{Brinch04}), for which we assumed
standard fractions of 70 and 30 percent to arise from dust in giant molecular clouds and in
the diffuse interstellar medium (\citealt{Ch00}). For S/N=30, the median star formation 
history recovered from the high-pass spectrum of this model does not differ in any systematic
way from that shown in Fig.~\ref{fig:Mock7Gyr}a, the discrepancies between the two cases
being less than 4 percent in all age bins.

\begin{figure}
\begin{minipage}{8.5cm}
\epsfig{file=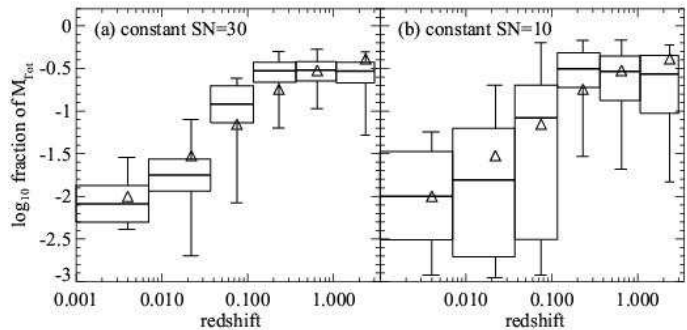,width=9cm}
\caption{Same as Fig.~\ref{fig:Mock7Gyr}, but for a galaxy with constant star 
formation rate.}
\label{fig:MockInfGyr}
\end{minipage}
\end{figure}

\begin{figure}
\begin{minipage}{8.5cm}
\epsfig{file=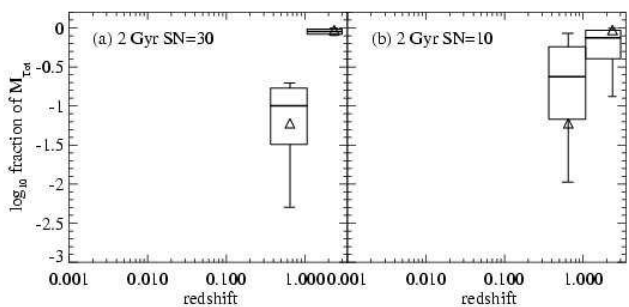,width=9cm}
\caption{Same as Fig.~\ref{fig:Mock7Gyr}, but for a galaxy with a star-formation time scale
of 2~Gyr.}
\label{fig:Mock2Gyr}
\end{minipage}
\end{figure}

Figs~\ref{fig:MockInfGyr} and \ref{fig:Mock2Gyr} show further examples of the recovery
of star formation histories from high-pass model galaxy spectra. Fig.~\ref{fig:MockInfGyr}
shows the results obtained for a galaxy with a constant star formation rate, while 
Fig.~\ref{fig:Mock2Gyr} shows those obtained for a galaxy with a star formation time scale
of 2~Gyr. In the latter case, we take the wavelength dependence of the noise to be that
observed for a typical early-type galaxy in the SDSS (Section~\ref{sec:EDR:DefGals}). For 
$\SNmed=30$, the method recovers quite well the star formation histories of both model 
galaxies (Figs~\ref{fig:MockInfGyr}a and \ref{fig:Mock2Gyr}a). The true mass fractions of
stars in all age bins are recovered within the 16--84 percent ($\pm1\sigma$) confidence 
ranges predicted by the analysis. We note that, for constant star formation rate
(Fig.~\ref{fig:MockInfGyr}a), the fraction of stars with ages around 1~Gyr is somewhat
better constrained than in the case of the 7~Gyr time-scale model
(Fig.~\ref{fig:Mock7Gyr}a). As before, reducing the signal-to-noise ratio from 30 to 10 
degrades the quality of the results, especially for the relative fractions of stars
in the oldest age bins. The solar metallicity of the models is recovered with 70 percent
confidence.

It is worth pausing here to emphasize the significance of the results presented in 
Figs~\ref{fig:Mock7GyrOldFormat}--\ref{fig:MockInfGyr}. These results are the first
to quantify the usefulness of the high-pass components of medium-resolution
galaxy spectra for constraining star formation histories. We have shown
that, for spectra with high-enough signal-to-noise ratio ($\SNmed\approx30$ per
pixel), the method outlined in Section~\ref{sec:Method} succeeds in recovering with
reasonable accuracy the star formation histories (and metallicities) of galaxies with 
a wide range of time scales of star formation. These results greatly extend the work of 
\citet{Ka03a}, who derived valuable but limited constraints on the recent star formation
histories of galaxies, based purely on the strengths of the 4000~{\AA} discontinuity and
H$\delta_{\rm A}$ stellar absorption line. We find that the analysis of the high-pass 
spectrum over  the whole wavelength range  from 3650 to 8500~{\AA} provides useful 
constraints on the {\em entire} star formation history of a galaxy, along with some
constraints on the stellar metallicity. 

A main advantage of our method over other approaches to extract star formation histories
from galaxy spectra is that it minimizes uncertainties linked to attenuation by dust
and the spectro-photometric calibration (Sections~\ref{sec:Method:Spec:Cover} and
\ref{sec:Method:ParSpace:Dust}). These uncertainties, which often must be included in 
a rudimentary way, have impaired most previous interpretation of galaxy spectra using 
low-resolution models (e.g., most recently, \citealt{Reic01, Pan03}). In this context, the
results presented in Figs~\ref{fig:Mock7GyrOldFormat}--\ref{fig:MockInfGyr} should be taken
as illustrative of the new type of constraints that can be set on the star formation 
histories of galaxies, thanks to the advent of medium-resolution population synthesis 
models. It is worth re-emphasizing  that, in the examples shown above, the simulated galaxy spectra
have the same solar heavy-element abundance ratios as the models used to interpret
them, while the abundance ratios of heavy elements are known to vary in external galaxies
(e.g., \citealt{Wor92,Eis03}). As mentioned in 
Section~\ref{sec:Method:Spec:Cover}, we exclude from our analysis the spectral features
with the strongest identified dependence on changes in the $\alpha$/Fe abundance ratio.
This should limit the uncertainties in the star formation histories and metallicities 
derived from observed, high-pass galaxy spectra (Section~\ref{sec:EDR}).

\subsection[]{The case of intermediate-age stars}
\label{sec:Testing:Signatures}

In the previous section, we noted the difficulty in recovering 
the mass fraction of stars with ages around 1~Gyr from the high-pass spectrum of a 13.5~Gyr-old 
model galaxy with a star-formation time scale of 7~Gyr (Fig.~\ref{fig:Mock7Gyr}a). This also
affected the constraints derived on the mass fractions of stars in the adjacent age bins.
We show here that this difficulty arises from a fundamental limitation of the
spectral interpretation of galaxies for which the optical signatures of 
intermediate-age stars (0.5--4~Gyr) are masked by those of younger and older stars.

\begin{figure}
\epsfig{file=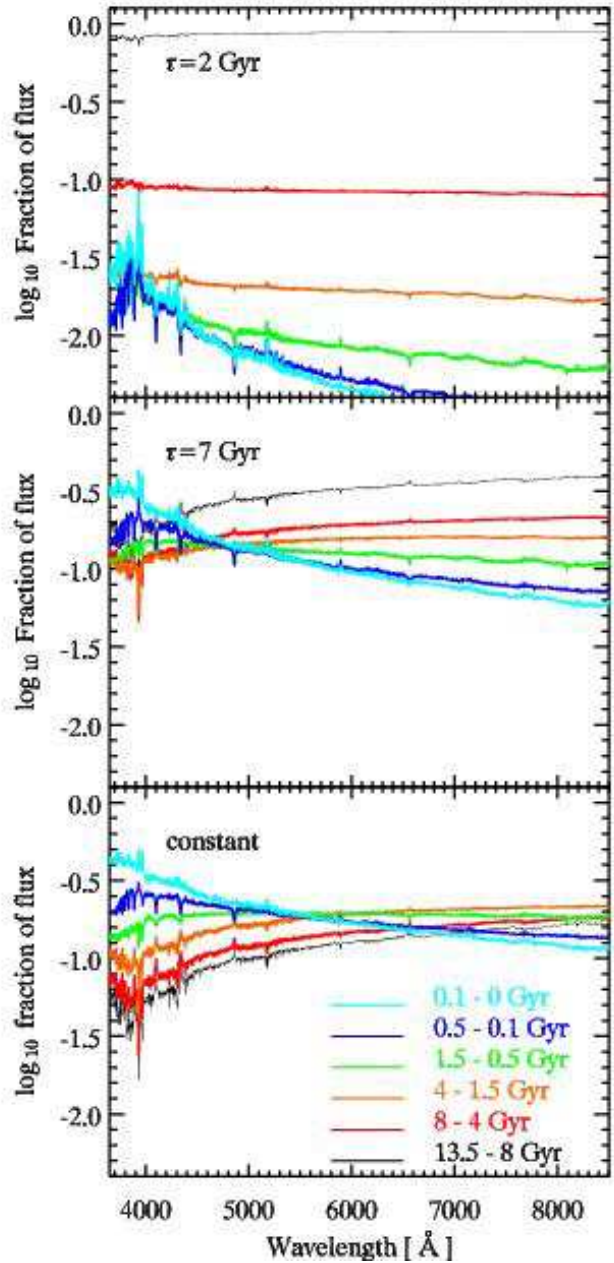,width=8.5cm} 		
\caption{Relative contributions by stars in different age bins to the integrated spectra 
of 13.5~Gyr-old model galaxies, for 3 different time scales of star formation, $\tau=2,$ 
7~Gyr and $\infty$ (constant star formation rate), and for solar metallicity. For
$\tau=7$~Gyr, stars with ages 0.5--4~Gyr contribute only little to the integrated light
at any wavelength. This makes the mass fraction of intermediate-age stars difficult to
constrain on the basis of the integrated spectra of galaxies with smoothly
declining star formation rates (see Fig.~\ref{fig:Mock7Gyr}).}
\label{fig:ContFractions}
\end{figure}

Fig.~\ref{fig:ContFractions} shows the relative contributions by stars in different age 
bins to the integrated spectra (including both the high-pass and low-pass components) of
the 3 model galaxies studied in Section~\ref{sec:Testing:MedianTime}, with star-formation
time scales $\tau=2,$ 7~Gyr and $\infty$ (constant star formation rate). For $\tau=2$~Gyr,
only stars older than 4~Gyr contribute significantly to the integrated light at any
wavelength. If the signal-to-noise ratio is high enough, the signatures of stars with 
ages 4--8 and 8--13.5~Gyr can be discerned in the integrated high-pass spectrum 
(Fig.~\ref{fig:Mock7Gyr}). For constant star formation rate, stars of any age always 
contribute to a major fraction of the integrated light in at least some wavelength range.
Young stars dominate at short wavelengths, while intermediate-age and old stars dominate
at long wavelengths. This makes the contributions by stars of all ages identifiable in the 
integrated high-pass spectrum of the galaxy (Fig.~\ref{fig:MockInfGyr}).

The case of the $\tau=7$~Gyr model is different. The few young stars in this model dominate
the integrated light at short wavelengths. However, the large number of old stars dominate
so much the emission at long wavelengths, that intermediate-age stars (0.5--4~Gyr) never
contribute more than a minor fraction of the integrated light at any wavelength. This is 
why the mass fraction of intermediate-age stars is difficult to constrain accurately from
the integrated high-pass spectrum of the galaxy, even for very high signal-to-noise ratio 
(Fig.~\ref{fig:Mock7Gyr}). This limitation does not affect the interpretation of only 
high-pass spectra. As Fig.~\ref{fig:ContFractions} shows, it is a fundamental limitation of
the spectral interpretation of galaxies with smoothly declining star formation rates, for
which the signatures of intermediate-age stars are masked by those of younger and older 
stars (an analogous limitation can arise when the spectrum is dominated by large amounts
of recent star formation; see Section~\ref{sec:EDR:SFHs} and Fig.~\ref{fig:SFH_Sp} below).
We note that $\tau\approx7$~Gyr is thought to be the volume-averaged time scale of 
star formation of the Universe (e.g., \citealt{Glaz03}; \citealt{Brinch04}). Hence, our
results imply that it may be difficult to constrain with precision the star formation 
history of the Universe at redshift $z\sim0.5$, based purely on the spectra of $z\sim0.1$
galaxies (Section~\ref{sec:Cosmic}).

\subsection[]{New insight into spectral indicators}
\label{sec:Testing:NewInd}

It is instructive to investigate the spectral features selected by the \MOPED\ algorithm to
constrain the star formation histories of galaxies from high-pass spectra. Given a fixed noise pattern, 
the weighting vectors constructed as outlined in Section~\ref{sec:Method:Overview} quantify in an optimal
way the sensitivity of the flux at each wavelength to the mass fractions of stars in various
age ranges. Thus, these vectors provide a means of identifying optimal spectral indicators
of the star formation histories of galaxies.

\begin{figure*}
\begin{minipage}{16cm}
\epsfig{file=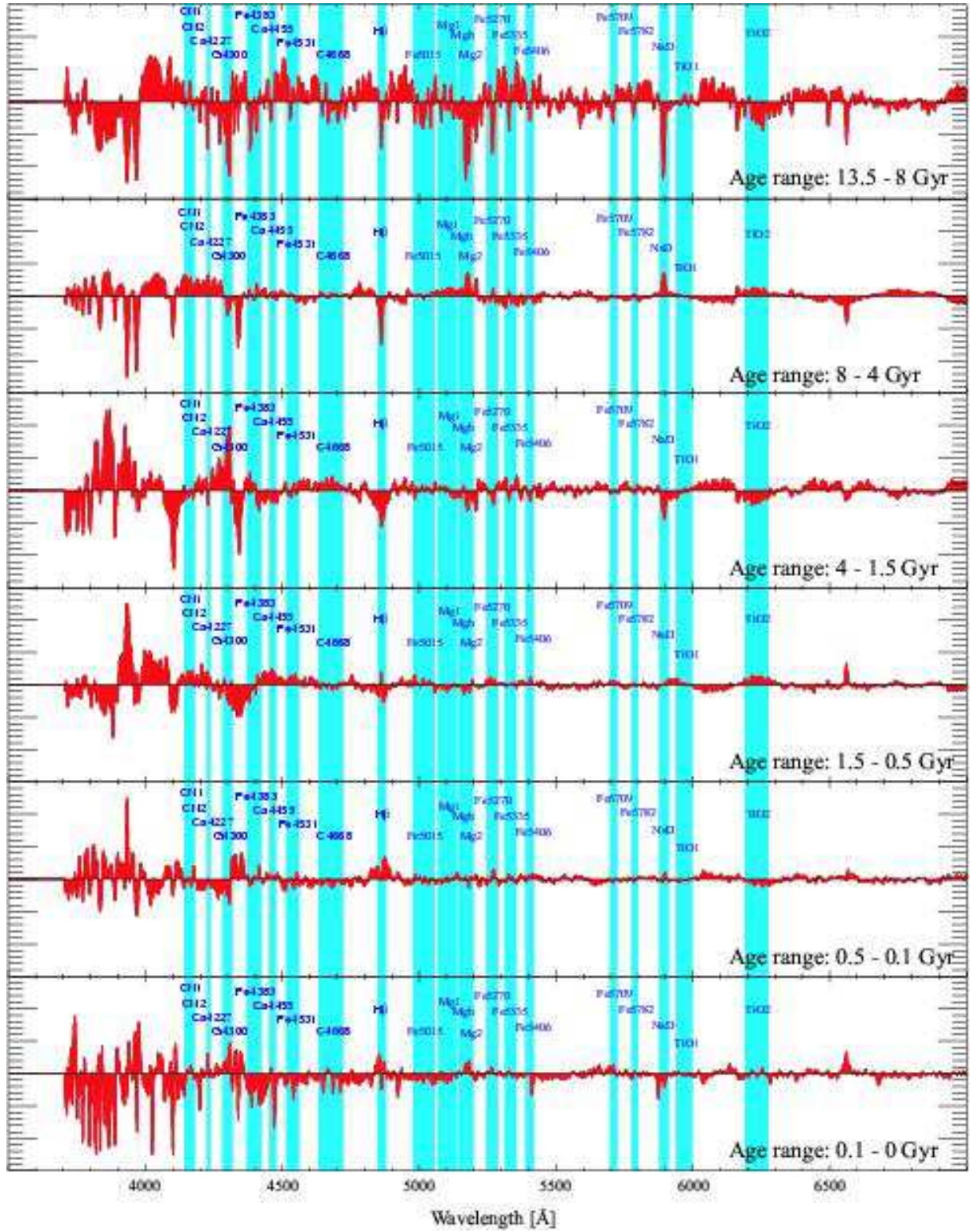,width=16cm}
\caption{Components of the \MOPED\ weighting vectors (red histogram) quantifying the sensitivity of the flux
at each wavelength to the mass fractions of stars in the 6 age bins used in the present 
analysis, as indicated in the legend. For illustrative purposes, the amplitudes have been
renormalized in each panel. Also shown for reference are the central band-passes of the 23 
spectral indices of the Lick system (blue bands). The predictions are for a stellar velocity
dispersion $\sigv=100\,\kms$, solar metallicity and a fixed signal-to-noise ratio 
per pixel  across the whole wavelength range from 3650 to 8500~{\AA}.}
\label{fig:ParAmpLick}
\end{minipage} 
\end{figure*}

Fig.~\ref{fig:ParAmpLick} shows the components of the weighting vectors quantifying the
sensitivity of the flux at each wavelength to the mass fractions of stars in the 6 age bins
used in our analysis (red histogram). In this example, we have adopted a stellar velocity 
dispersion $\sigv=100\,\kms$, solar metallicity and a fixed signal-to-noise ratio
across the whole wavelength range from 3650 to 8500~{\AA}. Also shown for reference in 
Fig.~\ref{fig:ParAmpLick} are the central band-passes of the 23 spectral indices of the 
Lick system (blue bands). Several of these, such as H$\delta_{\rm A}$, Ca4227, 
G4300, H$\gamma_{\rm A}$, H$\beta$, Ca4455 and NaD, coincide with spectral features 
identified by \MOPED\ to be most sensitive to the fractions of stars in various age ranges.
A full coincidence is not expected between the red histogram and the blue bands, because
many Lick indices are expected to be more sensitive to metallicity rather than age (e.g.,
\citealt{Wor94}).

\begin{figure*}
\begin{minipage}{16cm}
\epsfig{file=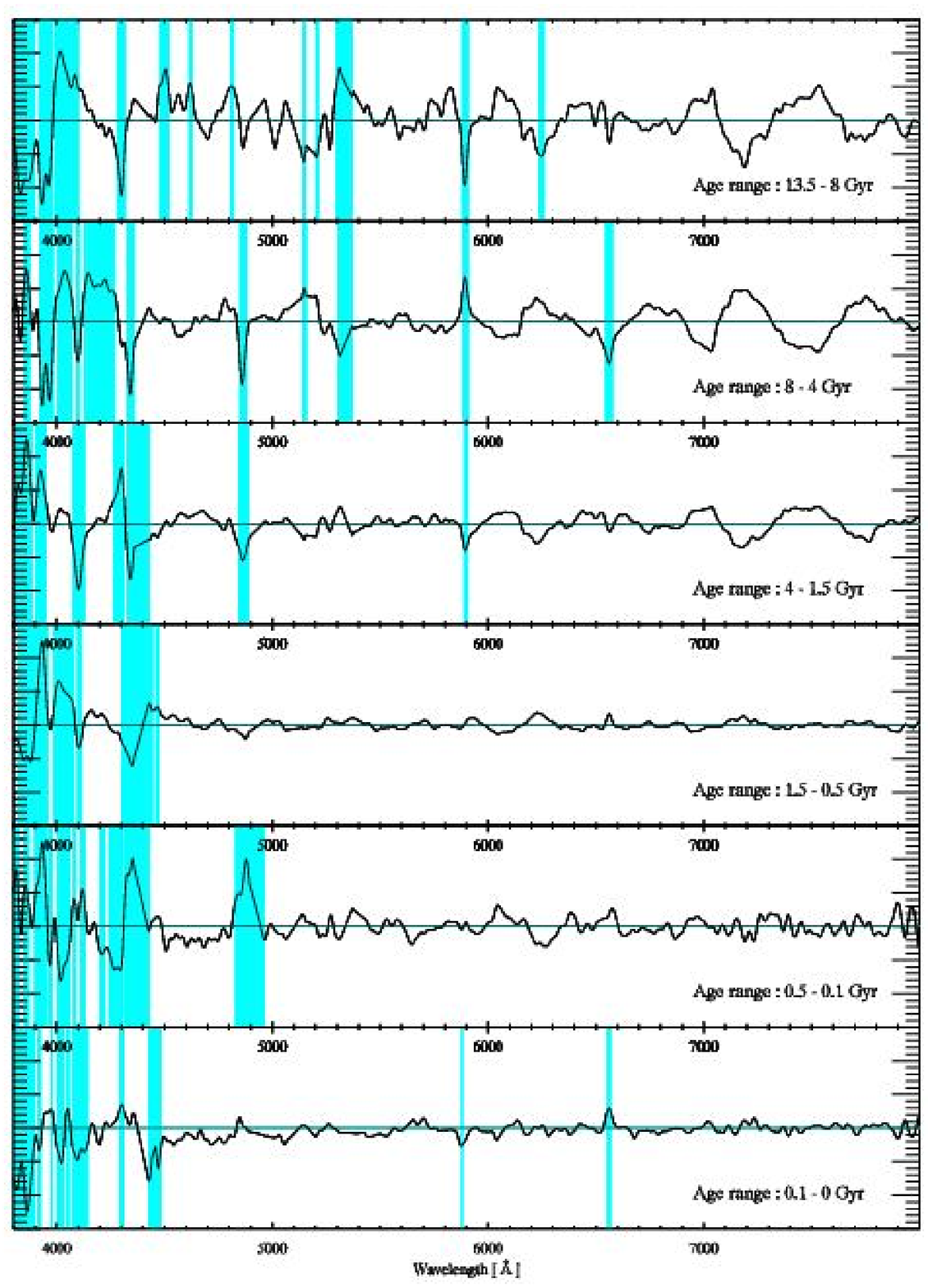,width=16cm}
\caption{Components of the \MOPED\ weighting vectors quantifying the sensitivity of the flux
at each wavelength to the mass fractions of stars in the 6 age bins used in the present
analysis (as indicated), after smoothing with a 
Gaussian kernel of  rms 10~{\AA} (solid
curve). For illustrative purposes, the amplitudes have been renormalized in each panel.
The blue bands mark spectral regions where the components of the smoothed vector exceed 
1.5 times the rms variation over the 3650--8500~{\AA} wavelength range. The prediction are
for a stellar velocity dispersion $\sigv=100\,\kms$, solar metallicity and a fixed 
signal-to-noise ratio across the whole wavelength range from 3650 to 8500~{\AA}.}
\label{fig:SmoothAmp}
\end{minipage} 
\end{figure*}

Several spectral regions identified by \MOPED\ to be highly sensitive to age in 
Fig.~\ref{fig:ParAmpLick} do not coincide with any Lick index. To highlight all the features
expected to reveal the presence of stars in various age ranges, we smooth each red 
histogram of Fig.~\ref{fig:ParAmpLick} using a Gaussian kernel of 
 rms 10~{\AA} and show the result as a solid curve in Fig.~\ref{fig:SmoothAmp}. For each age range, we
mark with blue bands the spectral regions where the components of the smoothed vector
exceed 1.5 times the rms variation over the 3650--8500~{\AA} wavelength range (we choose 
this level empirically to highlight only a moderate number of regions). Most of these 
regions appear to lie in the blue part of the optical spectrum; they could constitute the central band-passes 
of a set of optimal indicators of the presence of stars in various age ranges, which could be exploited without 
appealing to the full \MOPED\
analysis. In practice, only a few of the most sensitive indices should be selected to 
trace the presence of stars in each age range. We plan to explore this and the 
dependence of the spectral indicators on velocity dispersion and metallicity in more detail
in a forthcoming analysis.

\section[]{Application to the interpretation of observed galaxy spectra}
\label{sec:EDR}

In this section, we show examples of 
 derived star formation histories from 
observed, medium-resolution galaxy spectra using the method outlined 
in Section~\ref{sec:Method}. We first recall the main characteristics of the SDSS EDR
spectra we consider. We then show examples of derived star formation histories of 
early-type and late-type galaxies of comparable mass. Finally, we constrain the global
assembly of stellar mass in galaxies in different mass ranges in the Universe.

\subsection[]{SDSS galaxy spectra}
\label{sec:EDR:DefGals}

The observational sample we consider is the SDSS Early Data Release \citep{Stou02}. The
SDSS is obtaining $u$, $g$, $r$, $i$, and $z$ photometry of almost a quarter of the sky 
and spectra of at least 700,000 objects. The `main galaxy sample' of the EDR includes the
spectra of 32,269 galaxies with $r$-band Petrosian magnitudes in the range 14.50$<r<$17.77,
after correction for foreground Galactic extinction \citep{Strau02}. The median 
redshift is about 0.1. The SDSS spectra are acquired using 3-arcsecond diameter fibres that
are positioned as close as possible to the centres of the target galaxies. The spectra are
flux- and wavelength-calibrated, with 4096 pixels from 3800~{\AA} to 9200~{\AA} at
resolving power $\lambda/\Delta\lambda\approx1800$. This is similar to the resolution of 
our models in the wavelength range from 3650~{\AA} to 8500~{\AA}. We 
stress that, since we are interested in only the high-pass signal of the spectra, our results are not sensitive
to uncertainties in the spectro-photometric calibration of the SDSS EDR spectra 
(Section~\ref{sec:Method:Spec:Cover}).

We consider here the spectra of all galaxies with redshifts in the range $0.005<z<0.2$.
We use smear-corrected observations reduced through the {\tt spectro-1d} pipeline 
\citep{Subba02}, which we convert from vacuum to air for comparison with the models. We
remove sky (emission and absorption) lines. We also remove all detectable nebular emission
lines using the procedure described in \citet{Tre04} that is based on accurate fits to the
emission line-free regions of the spectra with model spectra broadened to the observed 
stellar velocity dispersion. Finally, we mask those regions of the spectra that are known
to strongly depend on changes in metal abundance ratios (Section~\ref{sec:Method:Spec:Cover}).

\subsection[]{The star formation histories of early-type and late-type galaxies}
\label{sec:EDR:SFHs}

To illustrate the constraints derived from high-pass spectra on the star formation
histories of galaxies, we differentiate between early-type and late-type galaxies.
We select the two types of galaxies on the basis of the `concentration parameter', 
defined as the ratio $C=R_{90}/R_{50}$ of the radii enclosing 90 and 50 percent of
the Petrosian $r$-band luminosity of a galaxy. This has been shown to correlate well
with `by eye' morphological classification \citep{Shi01,Str01}. \cite{Str01} propose
a cut at $C=2.6$ to separate early- from late-type galaxies. We classify conservatively
as `early type' all galaxies with $C>2.8$ and as `late type' all galaxies with $C<2.4$.
We exclude galaxies identified as hosts of active galactic nuclei (AGNs) by \citet{Ka03b}.
These various cuts reduce our sample to 3229 early-type and 9430 late-type galaxies with
redshifts $0.005<z<0.2$.

\begin{figure*}
\begin{minipage}{16cm}
\epsfig{file=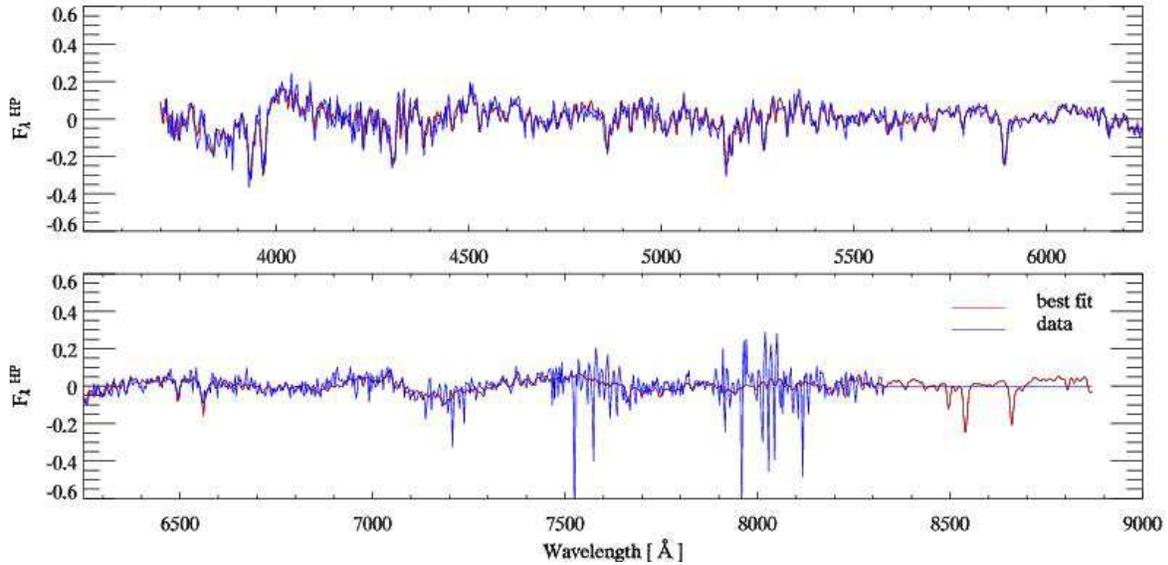,width=16cm}
\caption{Comparison of the high-pass spectrum of the early-type SDSS galaxy 
\#274--51913--392 (IAU: SDSS J103840.25+002251.6, in blue) with that of the model corresponding to the most likely star
formation history selected by the method described in Section~\ref{sec:Method} (in red).}
\label{fig:BestFitToEs}
\end{minipage}
\end{figure*}

\begin{figure*}
\begin{minipage}{16cm}
\epsfig{file=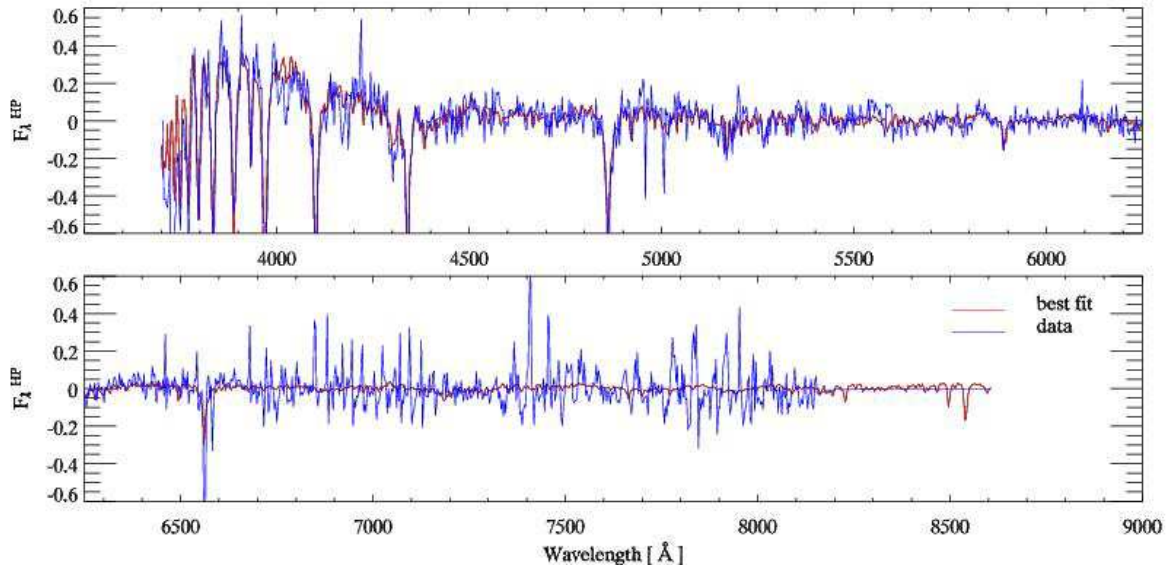,width=16cm}
\caption{Comparison of the high-pass spectrum of the  late-type  SDSS galaxy 
\#268--51633--202 (IAU: SDSS J095645.76+011448.7, in blue) with that of the model corresponding to the most likely star
formation history selected by the method described in Section~\ref{sec:Method} (in red).}
\label{fig:BestFitToSp}
\end{minipage}
\end{figure*}

We derive star formation histories for all these galaxies using the method described in 
Section~\ref{sec:Method}. In this procedure, we always adopt models with stellar velocity
dispersions close to the observed one and compute \MOPED\ weighting vectors specific to 
each galaxy, based on the observed noise spectrum (Section~\ref{sec:Method:Like:Fiducial}).
As an example, we compare in Figs~\ref{fig:BestFitToEs} and \ref{fig:BestFitToSp}
the observed high-pass spectra of an early-type galaxy 
(\#274--51913--392, IAU: SDSS J103840.25+002251.6) and a late-type galaxy 
(\#268--51633--202, IAU: SDSS J095645.76+011448.7) with the spectra of the models corresponding to the most likely
star formation histories selected by our analysis. The early-type galaxy has $C=2.95$,
a 4000\,{\AA} discontinuity $D4000=1.97$ (as computed using the narrow-band definition of
\citealt{Bal99}), a velocity dispersion $\sigv=127\,$\kms and a median signal-to-noise
ratio per pixel $\SNmed\approx27$. The late-type galaxy has $C=2.39$, $D4000=1.13$, $\sigv
\la75\,$\kms and $\SNmed\approx22$. There is good agreement between models and observations
for the main stellar (Balmer and metallic) absorption features in the spectra of both 
galaxies. 

\begin{figure}
\epsfig{file=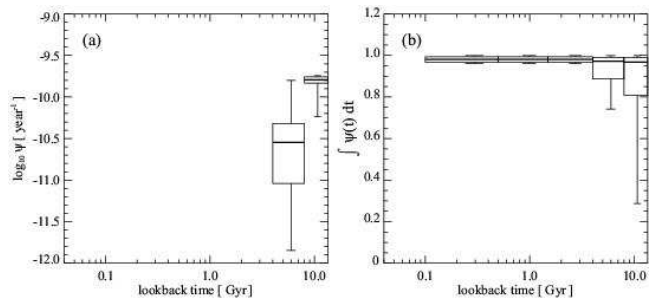,width=8.5cm}
\caption{Evolution of the star formation rate (left; normalized to $1\,M_\odot$ of stars
formed over all ages) and cumulative assembly of stellar mass (right) derived from the 
high-pass spectrum of the galaxy in Fig.~\ref{fig:BestFitToEs}. For each bin of look-back
time, the thick line, box and error bars indicate the median value, 16--84 and 2.5--97.5 
percentile ranges of the quantity plotted on the $y$-axis (these ranges indicate the 68 
and 95 percent confidence intervals in the recovered star formation history). Bins with 
no star formation are shown blank, as is the last bin of the cumulative assembly of stellar
mass (which is unity by construction).}
\label{fig:SFH_E}
\end{figure}

\begin{figure}
\epsfig{file=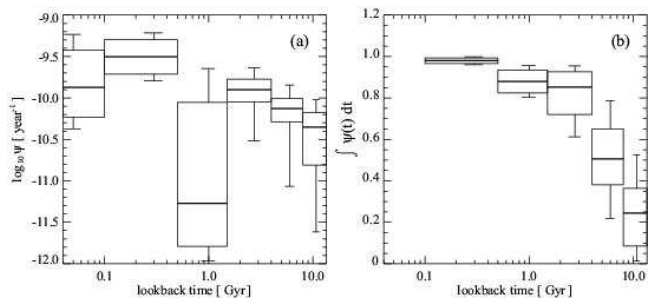,width=8.5cm}
\caption{Evolution of the star formation rate (left; normalized to $1\,M_\odot$ of stars
formed over all ages) and cumulative assembly of stellar mass (right) derived from the 
high-pass spectrum of the galaxy in Fig.~\ref{fig:BestFitToSp}. For each bin of look-back
time, the thick line, box and error bars indicate the median value, 16--84 and 2.5--97.5 
percentile ranges of the quantity plotted on the $y$-axis (these ranges indicate the 68
and 95 percent confidence intervals in the recovered star formation history). Bins with no
star formation are shown blank.} 
\label{fig:SFH_Sp}
\end{figure}

\begin{figure}
\epsfig{file=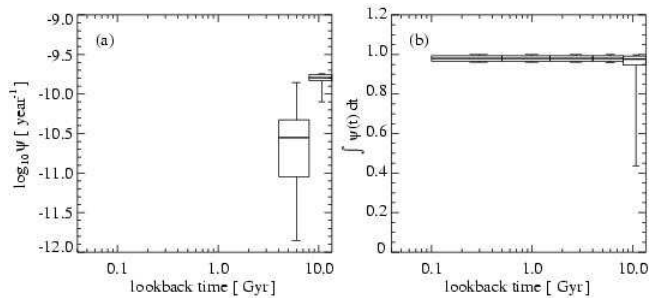,width=8.5cm} 		
\caption{Typical star formation history (left; normalized to $1\,M_\odot$ of stars
formed over all ages) and cumulative assembly of stellar mass (right) derived from the
high-pass spectra of early-type galaxies with stellar mass in the range $9.5<\log(M_\ast/
M_\odot)<11.5$ in the SDSS EDR. The results are based on the spectra of 659 galaxies
with concentration parameters $C>2.8$ (see text for detail). The boxes have the same
meaning as in Fig.~\ref{fig:SFH_E}.}
\label{fig:FracsE}
\end{figure}

\begin{figure}
\epsfig{file=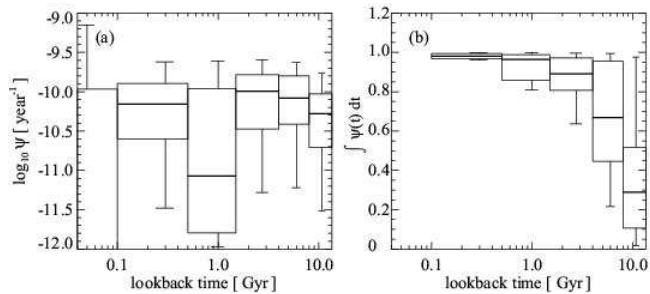,width=8.5cm} 		
\caption{Typical star formation history (left; normalized to $1\,M_\odot$ of stars
formed over all ages) and cumulative assembly of stellar mass (right) derived from
the high-pass spectra of late-type galaxies with stellar mass in the range $9.5<
\log(M_\ast/M_\odot)<11.5$ in the SDSS EDR. The results are based on the spectra 
of 385 galaxies with concentration parameters $C<2.4$ (see text for detail). The boxes
have the same meaning as in Fig.~\ref{fig:SFH_E}.}
\label{fig:FracsL}
\end{figure}

Figs~\ref{fig:SFH_E} and \ref{fig:SFH_Sp} show star formation histories derived from the
high-pass spectra of the early-type and late-type galaxies in Figs~\ref{fig:BestFitToEs}
and \ref{fig:BestFitToSp}, respectively. In each case, the left-hand panel shows the 
evolution of the star formation rate, while the right-hand panel shows the cumulative 
assembly of stellar mass. The star formation rates in Figs~\ref{fig:SFH_E} and 
\ref{fig:SFH_Sp} were computed by dividing the mass fraction of stars formed in each 
bin of look-back time by the duration of that bin. Thus, they are normalized to a total 
of $1\,M_\odot$ of stars formed over all ages. As in 
Figs~\ref{fig:Mock7Gyr}--\ref{fig:Mock2Gyr}, the thick line, box and error bars for each
bin of look-back time in Figs~\ref{fig:FracsE} and \ref{fig:FracsL} indicate the median
value, 16--84 and 2.5--97.5 percentile ranges of the quantity plotted on the $y$-axis
(these ranges indicate the 68 and 95 percent confidence intervals in the recovered star
formation history; the results of Section~\ref{sec:Testing:MedianTime} suggest that the
16--84 percentile range is typical of the accuracy to which our method can recover star 
formation histories from high-pass galaxy spectra). Fig.~\ref{fig:SFH_E} shows that most
of the stars in the early-type galaxy SDSS J103840.25+002251.6 are likely to have formed
over 8~Gyr ago, while star formation appears to have occurred much later in the late-type
galaxy SDSS J095645.76+011448.7. The significant recent star formation in this galaxy makes
the mass fraction of stars with ages around 1~Gyr difficult to constrain (see also 
Section~\ref{sec:Testing:Signatures}).

It is interesting to compare the {\em typical} star formation histories of early-type
and late-type galaxies of similar stellar mass. \citet{Ka03a} derived stellar masses
for the galaxies in the SDSS EDR, based on the observed strengths of $D4000$ and
of the H$\delta_A$ absorption line. We follow their procedure here and compute, for each 
galaxy in our sample, the median mass-to-light ratio in the observer-frame $i$ band that
is predicted by our analysis of the high-pass spectrum.\footnote{\citet{Ka03a} used the $z$
band in their analysis, which falls outside the spectral range of the medium-resolution 
models considered here for galaxies at redshifts $z\la0.2$ 
(Section~\ref{sec:Method:Spec:Cover}).} We assume that this ratio is the same for the whole
galaxy as it is in the region sampled by the fibre. We then compute the total stellar mass
$M_\ast$ from the observed $i$-band luminosity, after correcting the $i$-band flux for 
attenuation by dust and the contamination by emission lines, using the same procedure as 
outlined in \citet{Ka03a}. The galaxy stellar masses obtained in this way agree to within
50 percent with those derived by \citet{Ka03a}. For the SDSS-EDR sample, the distribution
of galaxy stellar masses peaks around $3\times10^{10}\,M_\odot$.

We focus on the typical star formation histories derived from the high-pass spectra of
early-type and late-type galaxies around the peak of the stellar mass distribution, i.e.
with masses in the range $9.5< \log(M_\ast/ M_\odot)<11.5$. One way to constrain these is
to stack the spectra of all galaxies of each type in this mass range and to derive the 
corresponding star formation history from the stacked spectrum. In practice, we first 
arrange the galaxies in 4 bins of velocity dispersion ($\sigv$), 5 bins of 4000-{\AA} 
discontinuity ($D4000$) and 2 bins of redshift ($z$). Then, within each 3-dimensional 
bin, we stack the (emission-line subtracted) spectra of all galaxies. In this procedure,
we normalize each spectrum to the total absolute magnitude derived from the observed
Petrosian $r$-band magnitude. We also weight each spectrum by a factor $1/\Vmax$, 
where $\Vmax$ is the largest volume within which the galaxy would pass the sample
selection criteria.\footnote{We compute $\Vmax$ using the effective area and the bright
(14.50 mag) and low (17.77 mag) magnitude cutoffs of the SDSS EDR. We do not
include any evolutionary corrections to $\Vmax$.} We derive the star formation history from the 
high-pass signal in the stacked spectrum in each bin of $\sigv$, $D4000$ and $z$. This 
star formation history corresponds to a final stellar mass density of $\Sigma_{i}(M_{\ast,i}
/V_{{\rm max},i})$. The typical star formation history of galaxies of a given type is 
therefore given by the $\Sigma_{i}(M_{\ast,i}/V_{{\rm max},i})$-weighted average of the
star formation histories of all bins for that type.

Figs~\ref{fig:FracsE} and \ref{fig:FracsL} show the typical star formation histories 
derived in this way from the high-pass spectra of 659 early-type and 385 late-type 
galaxies with stellar masses in the range $9.5<\log(M_\ast/M_\odot)<11.5$ (i.e., around
the peak of the stellar mass distribution of SDSS-EDR galaxies) in our sample. In each
case, the left-hand panel shows the typical, class-averaged evolution of the star 
formation rate, while the right-hand panel shows the cumulative assembly of stellar mass.
Figs~\ref{fig:FracsE} shows that most of the stars in the typical early-type galaxy in 
our sample are likely to have formed over 8~Gyr ago, although a small fraction of the 
total stellar mass can be accounted for by stars with ages down to 4~Gyr. This is consistent
with a wide range of constraints on the fraction of intermediate-age stars in early-type 
galaxies \citep{Jorg99,Tra00,Men01,Bel04}. We note that \citet{Reic01} used \MOPED\ to 
interpret the low-resolution spectra of 8 bright nearby E/S0 galaxies from the Kennicutt
(1992) atlas. They found that stars younger than 3~Gyr could account for over 30 percent
of the total stellar mass for 6 of these galaxies. This is not the case for the typical 
early-type galaxy in our sample.

Fig.~\ref{fig:FracsL} shows how the typical star formation history of a late-type galaxy
differs from that of an early-type galaxy in the same stellar-mass range. The typical 
late-type galaxy with $9.5<\log(M_\ast/M_\odot)<11.5$ appears to have formed stars at a
roughly constant or smoothly declining rate over the lifetime of the universe. The apparent
`drop' in the median star formation rate at look-back times around 1~Gyr in 
Fig~\ref{fig:FracsL}a is not statistically significant. It is a likely consequence of the 
difficulty of constraining the mass fraction of stars of this age for smoothly 
declining star formation histories (Section~\ref{sec:Testing:Signatures}). We note that the
average evolution of the star formation rate of a late-type galaxy in Fig.~\ref{fig:FracsL}a
is consistent with the star formation history of the Milky Way derived by \citet{Ro00a} 
from the analysis of 552 Galactic stars.

The results of Figs~\ref{fig:FracsE} and \ref{fig:FracsL} illustrate the accuracy to 
which our method can recover star formation histories from the observed high-pass 
spectra of individual galaxies in the SDSS EDR. The typical uncertainties in the derived
star formation histories are somewhat larger than those expected from the analysis of 
model galaxy spectra with similar signal-to-noise ratio ($\SNmed=30$) in 
Section~\ref{sec:Testing:MedianTime} (Figs~\ref{fig:Mock7Gyr}--\ref{fig:Mock2Gyr}). 
This arises in part from our simplifying assumption that all stars in the model galaxies
studied in Section~\ref{sec:Testing:MedianTime} had a fixed metallicity (corresponding to the
effective metallicity). The mixing of stars with different chemical compositions in real
galaxies makes the interpretation of the high-pass spectrum using single-metallicity 
models less precise. Also, our models have fixed, solar metal abundance ratios, whereas
the abundance ratios of heavy elements are expected to vary in external galaxies 
(Section~\ref{sec:Method:Spec:Cover}). The influence of this uncertainty on the present
study should be minimized by the fact that we have masked those features that are expected
to be most sensitive to changes in metal-abundance ratios in the SDSS galaxy spectra.

We  have also derived rough constraints on the metallicities of all galaxies in our sample by 
assigning them sub-solar ($0.4Z_\odot$), solar ($Z_\odot$) or supra-solar ($1.5Z_\odot$)
metallicity, as described in Section~\ref{sec:Method:Like:Sampl}. We have found that the
results compare relatively well to the more refined metallicities derived by \citet{Galla05}
for these galaxies, based on the analysis of selected Balmer and metallic lines with weak
dependence on metal-abundance ratios.

\subsection[]{Cosmic star formation history}
\label{sec:Cosmic}

We now use the high-pass spectra of SDSS galaxies to set constraints 
on the {\em global} star formation history of the Universe and its 
distribution among galaxies in different mass ranges. Here, we include all 
galaxies at redshifts $0.005<z<0.2$ in the SDSS EDR sample. We 
also include those galaxies identified as AGN hosts by \citet{Ka03b}. As shown 
by these authors, the presence of a (type 2) AGN does not alter significantly the
stellar absorption-line signatures in the host-galaxy spectrum. By analogy with 
our approach in Section~\ref{sec:EDR:SFHs}, we proceed by first stacking the
spectra of galaxies in different bins of velocity dispersion ($\sigv$; 4 bins), 
4000-{\AA} discontinuity ($D4000$; 3 bins) and redshift ($z$; 2 bins) to improve
the signal-to-noise ratio. As before, we normalize each individual spectrum to 
the total absolute magnitude derived from the observed Petrosian $r$-band magnitude
of the galaxy and apply a weight of $1/\Vmax$. Then, we derive the star formation
history from the high-pass signal in the stacked spectrum in each bin of $\sigv$, 
$D4000$ and $z$. This yields the probability density functions of the mass fractions
of stars formed in 6 age bins for this spectrum. This star formation history corresponds 
to a final stellar mass density of $\Sigma_{i}(M_{\ast,i}/\Vmax_{i})$, which we use
to transform stellar mass fractions into stellar mass densities. To obtain the 
global star formation history per unit comoving volume of the Universe, we then
simply combine the star formation histories over all bins of $\sigv$, $D4000$ and $z$.
In practice, this is obtained by combining the probability density functions of the
stellar mass densities formed in each redshift bin, following the precepts outlined
in appendix~A of \cite{Brinch04}.

\begin{figure}
\epsfig{file=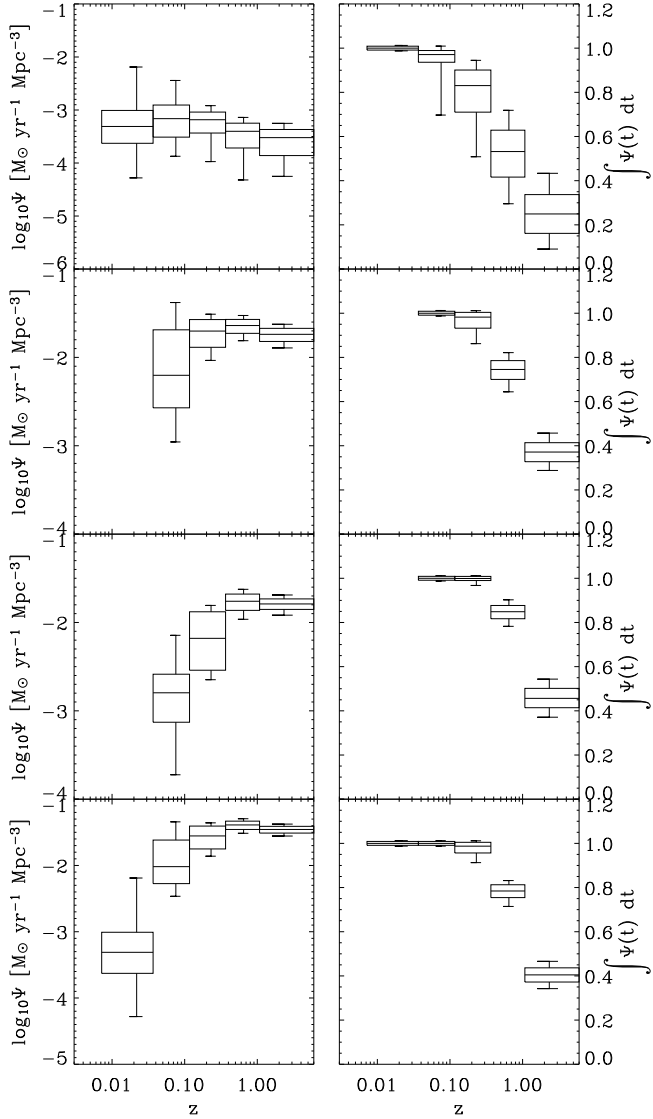,width=8.5cm}
\caption{Evolution of the star formation density (left) and cumulative
assembly of stellar mass (right) derived from the high-pass spectra 
of SDSS galaxies in different mass ranges at redshifts $0.005<z<0.2$. (a) For
galaxies with stellar masses in the range $7<\log(M_\ast/M_\odot)<9$. (b) For 
galaxies with stellar masses in the range $9<\log(M_\ast/M_\odot)<11$. (c) For 
galaxies with stellar masses in the range $11<\log(M_\ast/M_\odot)<13$. (d) For
galaxies with stellar masses in the full range $7<\log(M_\ast/M_\odot)<13$. For 
each bin of redshift, the thick line, box and error bars indicate the median 
value, 16--84 and 2.5--97.5 percentile ranges of the quantity plotted on the 
$y$-axis (these ranges indicate the 68 and 95 percent confidence intervals in the
recovered star formation history). Bins with no star formation are shown blank, and
the last bin of the cumulative assembly of stars is purposely omitted as it is unity
by construction.}
\label{fig:CosmicSFRByMass}
\end{figure}

Fig.~\ref{fig:CosmicSFRByMass} shows the evolution of the star formation density
(left) and cumulative assembly of stellar mass (right) as accounted for by 
galaxies in different mass ranges in a unit comoving volume of the Universe:
(a) $10^7$--$10^9 \,M_\odot$; (b) $10^9$--$10^{11} \,M_\odot$; (d)
$10^{11}-10^{13}\,M_\odot$; and (d) full mass range ($10^7$--$10^{13} \,M_\odot$).
The most striking result from this figure is the fact that the stars that
end up in massive galaxies today appear to have formed on average earlier than
those that end up in low-mass galaxies. This result was already obtained in
a different analysis by \citet{Hea04}, who used \MOPED\ to interpret the 
low-resolution spectra of 96,545 galaxies in the SDSS Data Release One (DR1). It
is consistent with the `down-sizing' scenario, in which the mass and luminosity of
the galaxies undergoing active star formation become progressively lower as the
Universe becomes older \citep{Cow96}. We note that both the present analysis and 
that of \citet{Hea04} rely on the spectral analysis of nearby galaxies. 
Hence neither analysis indicates whether the stars formed once the galaxies were
already assembled into systems with masses similar to the present-day ones. 

It is unlikely that the results of Fig.~\ref{fig:CosmicSFRByMass} be seriously 
affected by the magnitude limit of our sample. As shown by \cite{Brinch04}, at the
low-redshift limit of $z=0.005$, the $r=17.77$ magnitude limit of the spectroscopic
SDSS sample corresponds roughly to a stellar mass limit of just under $10^{8} 
M_{\odot}$ for a galaxy which would have formed all its stars 13.5~Gyr ago and faded
ever since. Most galaxies of the same mass containing younger stars would be brighter,
as attenuation by dust is not expected to be significant in low-mass galaxies (see
figure~6 of \citealt{Brinch04}). We find that restricting the stellar-mass range of
galaxies in Fig.~\ref{fig:CosmicSFRByMass}a to the more conservative completeness
range $8<\log(M_\ast/M_\odot)<9$ has a negligible effect on the derived star formation
history. Thus, the magnitude limit of the SDSS sample cannot account for the conclusion
from Fig.~\ref{fig:CosmicSFRByMass} that the stars in massive galaxies today formed on
average earlier than those in low-mass galaxies.

Aperture effects could potentially affect the results of Fig.~\ref{fig:CosmicSFRByMass}
because, at the low redshifts of the SDSS sample, the small-aperture SDSS fibres tend to
sample the inner, bulge-dominated regions of massive galaxies with bulge+disc morphologies.
\citet[][see also \citealt{Glaz03}]{Kew05} find that the influence of aperture effects on 
star-formation-rate estimates is significantly reduced when at least 20 percent of a 
galaxy's light is gathered in the fibre. They show that, for the average SDSS galaxy, this 
corresponds to restricting studies to redshifts larger than $z\sim 0.04$. Since most galaxies
more massive than $M_{\ast}\sim10^{9}
M_{\odot}$ in our sample lie at redshifts greater than this limit, aperture effects should have
only a weak influence on the results of Figs~\ref{fig:CosmicSFRByMass}b and c. Low-mass
galaxies in Fig.~\ref{fig:CosmicSFRByMass}a are not expected to be subject to strong aperture 
effects either, since their morphologies tend to be dominated by either a bulge or a disc, but
not both. This is supported by the finding by \citet{Brinch04} that low-mass SDSS galaxies 
show only weak colour gradients (we note that the star formation density in the lowest redshift
bin in Fig.~\ref{fig:CosmicSFRByMass}d, which is based on the spectra of only $\sim10$ 
nearby low-mass galaxies, is highly uncertain). Thus, we conclude that aperture effects 
should not strongly influence the results of Fig.~\ref{fig:CosmicSFRByMass}.

The evolution of the total star formation density of the Universe in
Fig.~\ref{fig:CosmicSFRByMass}d differs from that found by \citet{Hea04}
in that we do not find any statistically significant peak in the star formation
activity at redshift $z\la1$. This results in part from our more conservative
treatment of the uncertainties associated to the star formation histories 
derived with \MOPED\ (see Section~\ref{sec:Method:Like}). The relatively large
uncertainties in Fig.~\ref{fig:CosmicSFRByMass} also reflect the limitations of
current models for interpreting the high-pass spectra of observed galaxies (see
the discussion at the end of Section~\ref{sec:EDR:SFHs}). These uncertainties 
should be taken as very conservative, in the sense that they were constructed 
ignoring any possible correlation between the star formation histories of 
galaxies in different bins of velocity dispersion, 4000-{\AA}
discontinuity and redshift classes (see above). According to 
Fig.~\ref{fig:CosmicSFRByMass}d, about $40\pm5$ percent of all stars in the
Universe formed prior to redshift $z\approx1$, and the star formation density 
has since been declining smoothly. These results are consistent with those 
obtained by \citet{Bal02} from the analysis of the stacked, low-resolution 
spectra of 166,000 galaxies at redshifts $0.03<z<0.25$ extracted from the 
Two-degree Field Galaxy Redshift Survey (2dFGRS; \citealt{Colless01}). They 
conclude that the star formation density of the Universe must have peaked in 
the past ($z>0.6$) at a value at least three times the present-day one, while 
no more than 80 percent of all stars must have formed at redshifts $z>1$.


\section[]{Summary and Conclusions}
\label{sec:CCL}

We have presented a method for extracting star formation histories from medium-resolution
galaxy spectra, which is free of uncertainties arising from the spectro-photometric 
calibration and intrinsic attenuation by dust in the galaxies. The method focuses on the
interpretation of the high-pass signal in galaxy spectra, using a combination of the data
compression algorithm MOPED \citep{Hea00} with the medium-resolution population synthesis
code of \citet{Bru03}, which covers the wavelength range 3650--8500~{\AA} at a resolving
power of $\sim2000$. The principle of this method is to quantify the sensitivity of each pixel
in a high-pass galaxy spectrum to the mass fraction of stars formed in 6 age bins spanning
the age of the Universe. Given an observed spectrum, statistical estimates of the mass
fraction of stars formed in each age bin allow one to constrain the entire star formation
history of the galaxy. Our implementation of MOPED differs in several ways from previous
applications of this algorithm to the interpretation of low-resolution galaxy spectra 
\citep{Reic01,Pan03}. Overall, these differences should lead to more conservative estimates
of the errors in the derived star formation histories (Section~\ref{sec:Method:Like}). Our
approach does not require prior knowledge of the metallicity but assumes that all stars in
a given galaxy have the same metallicity, which should be regarded as the effective 
metallicity (i.e., the metallicity of the stars dominating the light).

We have tested in detail the ability of our method to recover star formation histories 
from model high-pass spectra of galaxies with different time-scales of star formation.
We find that the method can recover the full star formation histories of these models, 
provided that the median signal-to-noise ratio per pixel is large, i.e. 
$\SNmed>20$--30. Our method allows us to identify those spectral features that are most
sensitive to the presence of stars of a given age in the integrated spectrum of a galaxy.
We have exemplified how this could be used to define new `Lick-type' spectral indices
for studies of the star formation histories of galaxies (Section~\ref{sec:Testing:NewInd}).
The application of our method to model galaxy spectra also reveals a fundamental limitation
in the recovery of the star formation histories of galaxies for which the optical 
signatures of intermediate-age stars are masked by those of younger and older stars. One
of the consequences of this limitation is that the mass fraction of stars with ages around
1~Gyr is difficult to recover with accuracy in galaxies with smoothly declining star 
formation histories (i.e. with star formation time-scales of about 7~Gyr). Interestingly,
this type of star formation history is expected to be representative of the Universe as a 
whole \citep{Bal02}.

As an illustration, we have used our method to derive star formation histories from 
high-quality galaxy spectra in the SDSS EDR. To perform this analysis, we have first removed
emission lines from the observed spectra using the method outlined by \citet{Tre04}. We
have found that the star formation histories of morphologically identified early- and late-type
galaxies with stellar masses near the peak of the distribution in the local Universe, 
i.e. with $9.5<\log(M_\ast/M_\odot)<11.5$, are in agreement with what is commonly assumed 
about these systems. Early-type galaxies appear to have formed most of their stars over 8~Gyr
ago, although a small fraction of the total stellar mass of these galaxies may be accounted
for by stars with ages down to 4~Gyr. In contrast, late-type galaxies appear to have formed
stars at a roughly constant rate. We also examined the constraints set by the high-pass signal
in the stacked spectra of a complete magnitude-limited sample of 20,623 SDSS-EDR galaxies on
the global star formation history of the Universe and its distribution among galaxies in 
different mass ranges. Our results confirm that the stellar populations in the most massive
galaxies today appear to have formed on average earlier than those in the least massive ones
\citep{Hea04}. Our analysis does not reveal any statistically significant peak in the star
formation activity of the Universe at redshifts below unity.

At present, the main limitation in the interpretation of medium-resolution galaxy 
spectra is the  fixed, solar metal abundance ratios of the models at all metallicities 
while the abundance ratios of heavy elements are observed to vary in external galaxies
(e.g., \citealt{Wor92,Eis03}). To minimize the consequences of this limitation, we have 
masked those regions of galaxy spectra that are known to strongly depend to changes in 
metal abundance ratios (Section~\ref{sec:Method:Spec:Cover}). However, we suspect that at 
least part of the relatively large uncertainties affecting the star formation histories 
derived from the high-pass spectra of SDSS-EDR galaxies in Section~\ref{sec:EDR} may result
from this limitation. On the theoretical front, work is underway to develop the ability to
model the spectral evolution of galaxies with various metal abundances ratios at medium
resolution (e.g., Coelho et al., in preparation). It is worth noting that the (weak) 
constraints derived from our analysis on the metallicities of SDSS-EDR galaxies compare 
relatively well to the more refined ones derived for the same galaxies by \citet{Galla05} 
from the analysis of selected Balmer and metallic lines with weak dependence on metal-abundance
ratios (Section~\ref{sec:EDR:SFHs}).

The results obtained in this paper open the door to further refined interpretations
of galaxy spectra in terms of star formation history, metallicity and dust content. For
example, the low-pass signal in galaxy spectra, which is deliberately left aside in our
analysis, contains valuable information about intrinsic attenuation by dust. In 
Section~\ref{sec:Method:ParSpace:Dust}, we described how this information can be 
extracted from the low-pass signal, based on the constraints set by the high-pass signal
on the star formation history (this requires that the spectra be spectro-photometrically 
well calibrated). Another interesting development should come from the application of
our method to the interpretation of galaxy spectra gathered by high-redshift surveys, such
as the VIRMOS-VLT Deep Survey \citep{LeF01} and the Deep Extragalactic Evolutionary
Probe \citep{Dav03}. Such analyses will allow a more precise accounting of the star 
formation history of galaxies at the epoch $z\ga1$, which is poorly sampled by the analysis
of local galaxies. These various developments should lead us to a 
deeper understanding of how galaxies formed and evolved.



\section*{Acknowledgements}

We thank S. White, A. Boselli and B. Panter for useful discussions and A. Gallazzi
for comparing our metallicity estimates for SDSS galaxies to her more refined estimates 
in advance of publication. We also thank the referee, A. Heavens, for helpful comments.
H.M. thanks J. Silk for his encouragements and acknowledges support from PPARC. H.M. 
\& S.C. thank the Alexander von Humboldt Foundation, the Federal Ministry of Education 
and Research, and the Programme for Investment in the Future (ZIP) of the German Government
for their support. J.B. acknowledges the support of an ESA post-doctoral fellowship and an
FCT fellowship BPD/14398/2003.
	
Funding for the creation and distribution of the SDSS Archive has been provided by the Alfred
P. Sloan Foundation, the Participating Institutions, the National Aeronautics and Space 
Administration, the National Science Foundation, the U.S. Department of Energy, the Japanese
Monbukagakusho, and the Max Planck Society. The SDSS Web site is http://www.sdss.org/. 
 	
The SDSS is managed by the Astrophysical Research Consortium (ARC) for the Participating 
Institutions. The Participating Institutions are The University of Chicago, Fermilab, the
Institute for Advanced Study, the Japan Participation Group, The Johns Hopkins University,
Los Alamos National Laboratory, the Max-Planck-Institute for Astronomy (MPIA), the 
Max-Planck-Institute for Astrophysics (MPA), New Mexico State University, University of 
Pittsburgh, Princeton University, the United States Naval Observatory, and the University of
Washington.
 
\bsp
\bibliographystyle{mn2e}

\end{document}